\documentclass[twocolumn]{aastex61}
\usepackage{multirow}

\newcommand{\kmsend}{{km s$^{-1}$}}

\received{2018 August 13}
\revised{2019 January 21}
\accepted{2019 January 22}
\submitjournal{ApJ}

\shorttitle{Photometry of M85 GCs}
\shortauthors{Ko et al.}

\begin{document}

\title{A Wide-field Photometric Survey of Globular Clusters in the Peculiar Early-type Galaxy M85}

\correspondingauthor{Myung Gyoon Lee}
\email{mglee@astro.snu.ac.kr}

\author{Youkyung Ko}
\affiliation{Department of Astronomy, Peking University, Beijing 100871, People's Republic of China}
\affiliation{Kavli Institute for Astronomy and Astrophysics, Peking University, Beijing 100871, People's Republic of China}
\affil{Astronomy Program, Department of Physics and Astronomy, Seoul National University, 1 Gwanak-ro, Gwanak-gu, Seoul 08826, Republic of Korea}

\author{Myung Gyoon Lee}
\affiliation{Astronomy Program, Department of Physics and Astronomy, Seoul National University, 1 Gwanak-ro, Gwanak-gu, Seoul 08826, Republic of Korea}

\author{Hong Soo Park}
\affiliation{Korea Astronomy and Space Science Institute, 776 Daedeokdae-Ro, Yuseong-Gu, Daejeon 34055, Republic of Korea}

\author{Sungsoon Lim}
\affiliation{Herzberg Astronomy \& Astrophysics Research Centre, National Research Council of Canada, Victoria, BC V9E 2E7, Canada}

\author{Jubee Sohn}
\affiliation{Smithsonian Astrophysical Observatory, 60 Garden Street, Cambridge, MA 02138, USA}

\author{Narae Hwang}
\affiliation{Korea Astronomy and Space Science Institute, 776 Daedeokdae-Ro, Yuseong-Gu, Daejeon 34055, Korea}

\author{Byeong-Gon Park}
\affiliation{Korea Astronomy and Space Science Institute, 776 Daedeokdae-Ro, Yuseong-Gu, Daejeon 34055, Korea}



\begin{abstract}
We survey globular clusters (GCs) in M85 using $ugi$-band images of a $1 \arcdeg \times 1 \arcdeg$ field obtained with the MegaCam at the 3.6 m Canada-France-Hawaii Telescope. 
We identify 1318 GC candidates with 20.0 mag $< g_0 <$ 23.5 mag in the entire survey region.
Their radial number density profile is well fit by a S{\'e}rsic profile with $n$ = 2.58$^{+0.43}_{-0.33}$ and effective radius $R_{\rm e,GCS}$ = 4$\farcm$14 (= 22 kpc), showing that the candidates at $R < 20\arcmin$ are mostly genuine GCs in M85. 
We estimate the total number of GCs, $N$(total) = $1216^{+82}_{-50}$, and the specific frequency, $S_N = 1.41^{+0.10}_{-0.06}$.
The overall color distribution of the GCs in M85 is bimodal, but the GCs in the central region at $R < 2\arcmin$ do not show a bimodal distribution clearly.
The radial number density profile and surface number density map of the blue GCs (BGCs) show more extended structures than those of the red GCs (RGCs). 
The spatial distributions of both BGCs and RGCs are elongated, similar to that of the galaxy stellar light.
The number fraction of the RGCs in the central region  is much smaller compared to those in other early-type galaxies of similar luminosity.
The mean $(g-i)_0$ color of the RGCs in M85 is about 0.1 mag bluer than typical values for other Virgo early-type galaxies of similar luminosity, indicating that a significant fraction of the RGCs in M85 may be younger than typical GCs.
These results indicate that M85 might have undergone a major wet merger recently.
\end{abstract}




\section{Introduction}

According to the hierarchical galaxy formation scenario, massive early-type galaxies are formed by undergoing numerous mergers.
Globular clusters (GCs) are one of the excellent tools available to trace the assembly history of individual galaxies.
\added{The GCs in the Milky Way have a mean age of about 13 Gyr \citep{mar09,dot10,dsa11,van13}, indicating that the oldest GCs belong to the oldest populations in the universe.
Thousands of GCs are found in each massive early-type galaxy \citep{hbh17}.
The mean ages of the GCs in several massive early-type galaxies are estimated to be 7-13 Gyr from spectroscopy (e.g. Park et al. 2012). Thus, GCs keep the fossil record of the formation and evolution history of their host galaxies as well as themselves.}

In the past decade, there have been several observational studies with statistically meaningful samples about the physical properties of GCs associated with their host galaxies \citep{pen08, fvf09, geo10, vil10, chi11, liu11, cho12, hha13, hhh15, pl13, pot13, wan13, pas15, zar15, for16, for17, van17, amo18, fr18, lim18}.
One of the most prominent features of the GCs in massive early-type galaxies is their bimodal color distribution (see the references in \citet{bs06}).
\deleted{\citet{pen06a} suggested that this GC color bimodality is more obvious in bright galaxies than in fainter ones based on the analysis of the GCs in 100 early-type galaxies from the ACS Virgo cluster Survey (ACSVCS; C{\^o}t{\'e} et al. 2004).}
The GC color bimodality indicates that there are two 
distinguishable populations residing in the same galaxy.
The color of the old GCs can be a proxy for their metallicity, so the blue GCs (BGCs) and red GCs (RGCs) are thought to be metal-poor and metal-rich populations, respectively  \added{(e.g., Brodie et al. 2012)}.
These two GC subpopulations show significant differences in their radial extents and two-dimensional structures.
The radial distributions of the BGCs in several massive early-type galaxies are found to be more extended to the outer region of galaxies than those of the RGCs \citep{kis97,lkg98,str11,fpo12}.
Furthermore, the RGC systems in massive early-type galaxies are more elongated along their host galaxy stellar light than the BGC systems, although both GC systems are well aligned with the major axis of their host galaxies \citep{pl13,		 wan13}.
All of these studies show that the RGCs are much more correlated with the stellar light in their host galaxies compared with the BGCs.
These results suggest that massive early-type galaxies may have dual halos, a metal-poor halo and a metal-rich halo \citep{pl13}.

There have been several formation scenarios proposed to explain the presence of these two subpopulations of GCs in massive early-type galaxies.
\citet{az92} and \citet{za93} suggested that elliptical galaxies are formed by gas-rich major mergers between disk galaxies.
In this model, metal-poor GCs originated from the progenitor spirals and metal-rich GCs were formed during the merging epoch.
\citet{fbg97} suggested a multiphase collapse scenario where metal-poor GCs were formed during the early stage of galaxy formation at high redshift and metal-rich GCs were formed during the second collapse phase after a few gigayears.
\citet{cot98} suggested an accretion scenario where metal-rich GCs were formed in a massive seed galaxy and metal-poor GCs were accreted from nearby low-mass galaxies.
\citet{lee10} proposed a mixed model that includes the key ingredients of the previous models described above.
In this model, metal-poor GCs were mainly formed in low-mass dwarf galaxies and metal-rich GCs were formed in massive galaxies in-situ or in dissipative merging galaxies after metal-poor GCs.
Massive galaxies grow via dissipationless mergers with surrounding dwarf galaxies.
Consequently, metal-poor GCs in the outer part of massive galaxies mainly originated from dwarf galaxies via accretion.

All of the above explanations about GC formation assume that both BGCs and RGCs were formed at an early epoch. 
However, every galaxy has a different merging history, and some galaxies might have a dissipative merger much after both BGCs and RGCs were formed.
If so, some galaxies may host an intermediate-age GC population that was formed during the gas-rich merging events that occurred at a later stage.
In this study, we investigate the GC system in M85, one of the nearest massive early-type galaxies, \replaced{M85}{which} is known to host intermediate-age GCs. \deleted{but rarely studied.}

M85 (NGC 4382, VCC 798) is the fifth-brightest early-type galaxy in the Virgo Cluster, classified as S0$_{1}$(3)pec in \citet{bst85}, SA0$^{+}$pec in \citet{dev91}, and E2 in \citet{kor09}.
It is located in the northernmost region of the Virgo Cluster, far from the main body of the Virgo Cluster, where bright early-type galaxies are located.

\added{M85 shows peculiar features not only in its structure and kinematics but also in its GC system.}
It shows several merger-induced features such as distorted isophotes \citep{bur79}, shells, ripples \citep{ss88}, boxy isophotes \citep{fer06}, and a kinematically decoupled core \citep{mcd04}.

In addition, the GCs in M85 do not show a bimodal color distribution clearly, unlike the GCs in other bright galaxies in Virgo.
Based on the data for the GCs in 100 early-type galaxies from the ACS Virgo Cluster Survey (ACSVCS; C{\^o}t{\'e} et al. 2004),
\citet{pen06a} suggested that M85 is the best candidate that may have a trimodal GC color distribution.
This trimodal color distribution might originate from the presence of an intermediate-age GC population.
\added{\citet{chi11} investigated the age distribution of the GCs in 14 early-type galaxies, including M85, based on $gzK$-band photometry. They found that M85 hosts  GCs younger than those in other elliptical galaxies.}
In addition, \citet{tra14} investigated the GCs in the northwestern region of M85 based on optical photometry from the ACSVCS and $K_{\rm s}$-band photometry taken with the Near InfraRed Imager and spectrograph (NIRI) attached at the Gemini North telescope.
They suggested that about 85\% of the observed GCs were formed about 1.8 Gyr ago.
Recently, \citet{ko18} presented a Gemini/GMOS spectroscopic study of the GCs in the central region of M85, and they showed that there is a $\sim$4 Gyr old GC population in addition to the old one.

Besides, \citet{pen08} presented the specific frequency of the GCs in the ACSVCS galaxies. They derived the specific frequency of the M85 GCs as $S_{\rm N}$ = 1.29 $\pm$ 0.21 using the radial number density profile of the GCs based on the ACSVCS and WFPC2 imaging data and the GC luminosity function (GCLF) from \citet{jor07}.
However, all of these previous studies of the GCs covered only a small extent of M85 (less than $\sim 5\arcmin \times 5\arcmin$) because of the small field of view of the instruments used in their studies.

In this study, we present the first wide-field photometric survey of the GCs in M85 covering a $1 \arcdeg \times 1 \arcdeg$ field, using the MegaCam attached at the 3.6 m Canada-France-Hawaii Telescope (CFHT).
This paper is organized as follows. 
We briefly describe our observations and data reduction in Section 2.
In Section 3, we identify the GC candidates in the survey region and investigate their color distribution. 
Then, we present spatial distributions of the GC subpopulations.
The main results are discussed in Section 4 and summarized in the Section 5.
We adopted a distance to M85 of 17.9 Mpc based on surface brightness fluctuation measurements \citep{mei07}. One arcminute corresponds to 5.21 kpc at the distance to M85.

\section{Observations and Data Reduction}

	\begin{figure}[hbt]
\includegraphics[width=\columnwidth]{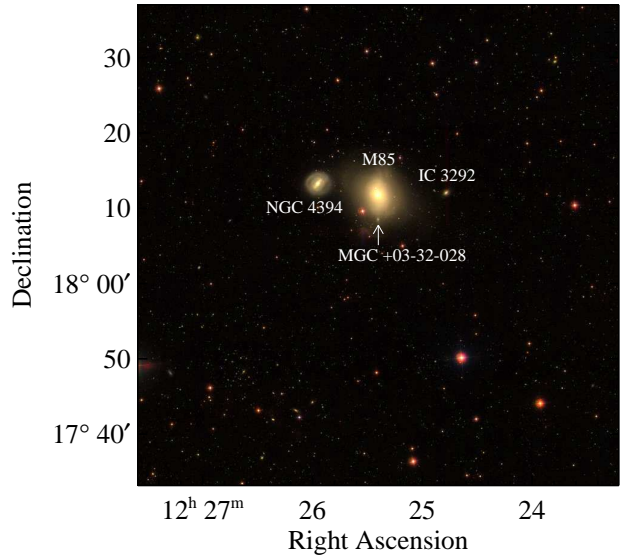}
\epsscale{1}
\caption{SDSS color image of the $1^{\circ} \times 1^{\circ}$ field for M85.
Note that M85 and its three nearby galaxies compose a small group, and they are located in the very low density region.
\label{fig:finder}}
	\end{figure}

	\subsection{Observation}

We carried out a wide-field photometric survey of GCs in M85 using MegaCam \citep{bou03} mounted on the CFHT (program ID: 14AK06; PI: Myung Gyoon Lee) during 2014 May and July.
The CFHT/MegaCam {\bf is} composed of 9 $\times$ 4 CCDs with a pixel scale of 0$\farcs$187 and covers a $1 \arcdeg \times 1 \arcdeg$ field.

{\bf Figure \ref{fig:finder}} shows the observation field.\deleted{covering $1\arcdeg \times 1 \arcdeg$ field around M85.}
M85 is surrounded by the barred spiral galaxy NGC 4394 and two dwarf galaxies IC 3292 and MGC +03-32-028.
The radial velocities of M85, NGC 4394, IC 3292, and MGC +03-32-028 are 772, 1001, 696, and 1228 \kmsend, respectively, from the Sloan Digital Sky Survey (SDSS) DR7 catalog \citep{aba09}.
These four galaxies compose a small group.
There are seen no other possible members of this group as seen in {\bf Figure \ref{fig:finder}}.
Thus, the M85 group is located in a very low density region.

We used $ugi$ filters because they provide two color combinations that have been extensively used for selecting GC candidates \added{(e.g. Cantiello et al. 2018; Ordenes-Brice{\~n}o et al. 2018)}.
\replaced{The exposures for $u$, $g$, and $i$ filters were taken in five times of 720 s, five times of 463 s, and seven times of 575 s, respectively.}{Exposure times were 5 $\times$ 720 s for $u$, 5 $\times$ 463 s for $g$, and 7 $\times$ 575 s for $i$.}
{\bf Table \ref{tab:log}} shows the observation log.
Mean values of the seeing were 0$\farcs$79 for $u$, 1$\farcs$06 for $g$, and 0$\farcs$60 for $i$.
We used a small dithering pattern for individual exposures for each filter.
The small gaps between the individual CCDs were filled by this dithering pattern, while the horizontal large gaps between the first and second rows and between the third and last rows of the CCDs were filled only partially.
The basic data reduction, including the standard bias and flat-field corrections, was performed with the `Elixir' pipeline \citep{mc04}.

\begin{deluxetable}{c c c c c c}
\tabletypesize{\footnotesize}
\tablecaption{Observation Log \label{tab:log}}
\tablewidth{\columnwidth}
\tablehead{\colhead{Filter} & \colhead{$N$(exp)} & \colhead{$T$(exp)} & \colhead{Air Mass} & \colhead{Seeing} & \colhead{Date (UT)} \\
\colhead{} & \colhead{} & \colhead{(s)} & \colhead{} & \colhead{(arcsec)}}
\startdata
$u$ & 5 & 720 & 1.018 & 0.79 & 2014 May 28 \\
$g$ & 5 & 463 & 1.039 & 1.06 & 2014 May 26 \\
$i$ & 5 & 575 & 1.004 & 0.60 & 2014 May 29 \\
$i$ & 2 & 575 & 1.538 & 0.56 & 2014 Jul 24 \\
\enddata
\end{deluxetable}

	\subsection{Photometry and Standard Calibration}

We performed astrometry for $ugi$-band images with {\sc SCAMP} \citep{ber06} using the point-source catalog from SDSS DR7 \citep{aba09}.
The images for each filter were stacked with {\sc SWARP} \citep{ber02} after the galaxy light of M85 and the nearby spiral galaxy NGC 4394 were subtracted with the IRAF/ELLIPSE task.
We removed the sky background from each image and set the sky level as zero.

We used SExtractor \citep{ba96} on the $ugi$ band stacked images for source detection and photometry.
We detected sources in the $i$-band image with a threshold of 3$\sigma$ and performed aperture photometry with aperture diameters of 8 pixels on the $ugi$ images with the dual-image mode.
The standard calibration of the instrumental magnitudes was done with SDSS DR12 $ugi$ point-spread function (PSF) photometry \citep{ala15} after the SDSS magnitudes were transformed to the CFHT/MegaCam magnitude system\footnote{http://www.cadc-ccda.hia-iha.nrc-cnrc.gc.ca/en/megapipe/docs/filt.html}.
We divided the survey region into 20 subregions for the $u$-band image and 36 subregions for the $gi$-band images and applied the calibration differently depending on the location of each source.
\added{The number of subregions was selected to include more than 20 standards in each subregion.
The zero-points for subregions range from 31.63 to 31.76 mag in $u$ band, from 32.15 to 32.30 mag in $g$ band, and 32.41 to 32.51 mag in $i$ band with standard deviations less than 0.03 mag.}
Standard calibration errors are, on average, 0.04, 0.02, and 0.02 mag for $u$-, $g$-, and $i$-band photometry, respectively.
Hereafter, all photometry results of this study are based on the AB magnitudes in the CFHT/MegaCam filter system.

	\begin{figure}[t]
\includegraphics[trim={1cm 0.5cm 1cm 0},clip,width=\columnwidth]{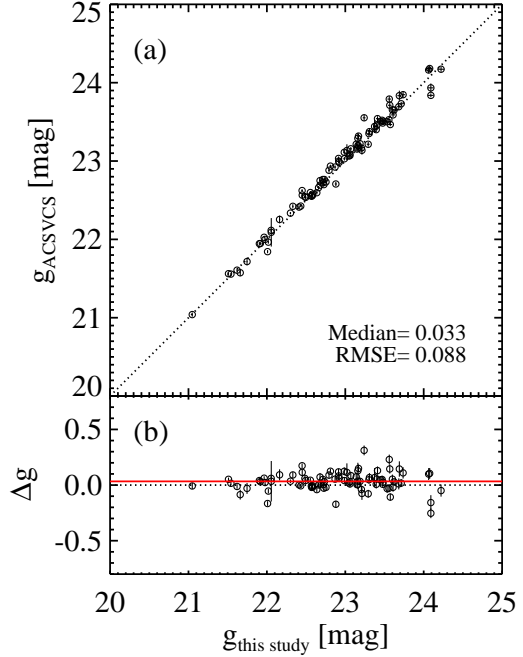}
\epsscale{1}
\caption{(a) Comparison of the $g$-band magnitudes measured in this study with those in the ACSVCS \citep{jor09}.
For comparison, both magnitudes are converted to SDSS $g$-band AB magnitudes without foreground reddening correction.
The horizontal and vertical error bars represent the measurement errors of the magnitudes.
The dotted line denotes the one-to-one relation.
(b) The $g$-band magnitude differences ($g_{\rm ACSVCS} - g_{\rm this~study}$) vs. $g$-band magnitudes in this study.
The dotted and solid lines indicate the zero level and the median difference, respectively.
\label{fig:comp_mag}}
	\end{figure}
	
{\bf Figure \ref{fig:comp_mag}} shows the comparison of $g$-band magnitudes derived in this study with those in the GC catalog of the ACSVCS \citep{jor09}.
The F475W magnitudes are the same as the dereddened SDSS $g$-band AB magnitudes.
For comparison, we converted the magnitudes from both this study and \citet{jor09} to the SDSS $g$-band AB magnitudes without foreground reddening correction.
There are 89 common sources between this study and \citet{jor09}, and the $g$-band magnitudes in both studies agree well.
The median value of the difference between the two is $\Delta g ({\rm ACSVCS} - {\rm this~~ study})=0.033$ mag, and its rms and mean error are 0.088 mag and 0.009 mag, respectively.

\added{
	\subsection{Completeness test}
	
We estimated the photometric completeness for GC candidates using an artificial source experiment with $g$-band images.
We divided the survey region into 36 subregions and added 100 artificial sources with an FWHM of 6 pixels ($\sim 1\farcs1$) into each subregion using the IRAF/ADDSTAR task.
These sources have magnitude ranges of 16 mag $< g_0 <$ 26 mag and GC candidate colors.
We repeated this procedure 100 times, and the total number of added artificial sources in 100 test images is 360,000.
Photometry was done for the test images with the same procedure as described in Section 2.2.
The completeness was derived from the number ratio of the recovered sources to the added sources.

	\begin{figure}[t]
\includegraphics[trim={1cm 0.5cm 0 0},clip,width=\columnwidth]{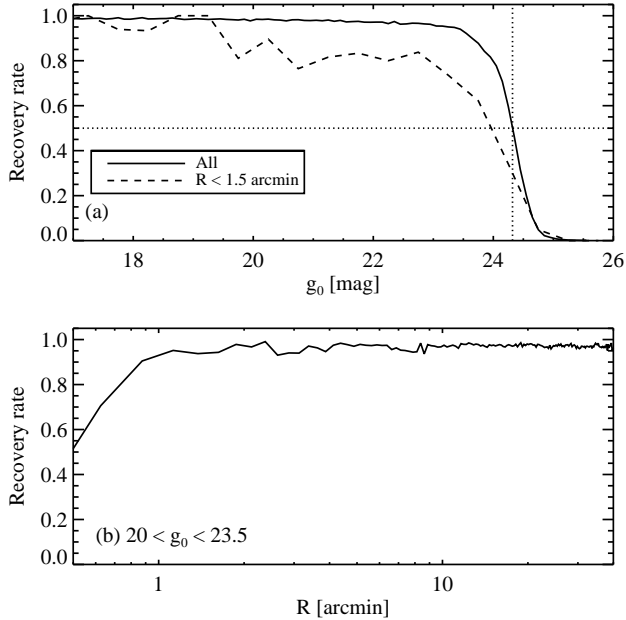}
\epsscale{1}
\caption{\added{(a) Completeness fraction for point sources (with FWHM = 6 pixels) as a function of $g_0$ magnitude.
The solid and dashed lines correspond to the entire survey region and central region ($R < 1\farcm5$), respectively.
(b) Completeness fraction for point sources with 20 mag $< g_0 <$ 23.5 mag as a function of galactocentric distance from M85.}
\label{fig:completeness}}
	\end{figure}

{\bf Figure \ref{fig:completeness}(a)} shows the completeness curve as a function of $g_0$ magnitude.
The recovery rate of all added sources is almost 100\% for bright magnitudes and steeply decreases at fainter magnitudes, reaching 50\% at $g_0 \sim 24.32$ mag.
The completeness of the central region ($R < 1\farcm5$) is lower than that of the entire region on average.
{\bf Figure \ref{fig:completeness}(b)} shows the radial variation of the completeness for the sources with 20 mag $< g_0 <$ 23.5 mag.
This magnitude range is the same as that used to select GC candidates (see Section 3.1).
The photometric completeness is higher than 90\% at $R > 0\farcm9$. It decreases as $R$ decreases at $R < 0\farcm9$ and drops to 65\% at $R = 0\farcm65$.
}

\section{Results}

	\subsection{GC Candidate Selection}

We selected GC candidates using three criteria: magnitude concentration parameter, color-color diagram, and magnitude ranges. Each is described in the following.

First, we selected GC candidates among the point-like and slightly extended sources detected in the survey region.
Most GCs are expected to be point-like or slightly extended in MegaCam images at the distance of the Virgo Cluster.
\replaced{We select those sources based on the magnitude concentration index $C$ defined as the $i$-band magnitude differences between 4 pixel and 8 pixel diameter aperture photometry, following the methods applied to the CFHT images of the Virgo cluster by \citet{dur14}.
\citet{dur14} presented a large-scale distribution of GCs in the Virgo cluster, based on photometry from Next Generation Virgo cluster Survey (NGVS; Ferrarese et al. 2012).}
{The sources in the images are expected to be marginally resolved if their sizes are at least 0.1 $\times$ FWHM(PSF) \citep{har09}. The seeing value of the $i$-band images in this study is $\sim$0$\farcs$6, corresponding to $\sim$52 pc at the distance of M85. Therefore, we can marginally resolve the sources with sizes of $\sim$5 pc (with effective radii $>$ 2.5 pc). \citet{dur14} confirmed that the magnitude concentration index $C$, defined as the $i$-band magnitude differences between 4-pixel and 8-pixel aperture photometry, is effective at distinguishing these resolved sources in the CFHT/MegaCam images. They presented a large-scale distribution of GCs in the Virgo Cluster based on photometry from the Next Generation Virgo Cluster Survey (NGVS; Ferrarese et al. 2012). We adopted the same concentration index criterion as in \citet{dur14} because the $i$-band seeing in this study is {\bf the} same as theirs. We selected nonresolved sources with $-0.1 < C < 0.1$ and marginally resolved sources with $0.1 < C < 0.15$.}

\begin{figure}[t]
\includegraphics[trim={0 0.5cm 0 0},clip,width=\columnwidth]{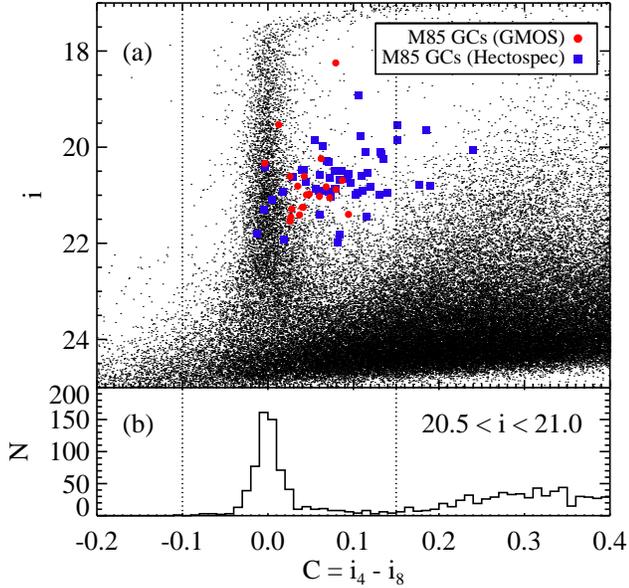}
\epsscale{1}
\caption{(a) $i$-band magnitudes vs. $i$-band concentration index ($C = i_4 - i_8$) for the sources detected in the survey region.
The stellar sequence is clearly seen at $C \sim 0$.
The filled circles and squares represent the GCs in M85 confirmed by the spectroscopic observations from Gemini/GMOS \citep{ko18} and MMT/Hectospec (Ko, Y. et al. 2019, in preparation), respectively.
Note that the confirmed GCs are slightly more extended than the stars in the $i$-band image taken in this study.
The dotted lines represent the concentration index criteria for selecting GC candidates, as used in \citet{dur14}.
(b) Concentration index distribution of the sources with 20.5 mag $< i <$ 21.0 mag.
\label{fig:c_index}}
	\end{figure}

{\bf Figure \ref{fig:c_index}(a)} shows the $i$-band magnitude versus concentration index diagram.
The point-source sequence is clearly shown at $C \sim 0$, while extended sources have much larger concentration indices.
The histogram of the $C$ values for the bright sources with 20.5 mag $< i <$ 21.0 mag shows a strong narrow component with a peak at $C \sim 0$ and a much broader and weaker component with a peak at $C \sim 0.3$ (see {\bf Figure \ref{fig:c_index}(b)}). 
The narrow component is mainly due to point sources, while the broad component is due to extended sources.
However, it is noted that the right side of the narrow component shows a long tail.
This is due to slightly extended sources, \deleted{(called compact sources hereafter)} which include slightly resolved GCs in M85.
We used the concentration index criterion, $-0.10 < C < 0.15$, for selecting point-like and \replaced{compact}{slightly extended} sources.
\citet{dur14} adopted the same selection criterion for Virgo GC selection.

In {\bf Figure \ref{fig:c_index}(a)} we plot known GCs in M85 that were spectroscopically confirmed based on Gemini/GMOS \citep{ko18} and MMT/Hectospec observations (Ko, Y. et al. 2019, in preparation).
Their $i$-band magnitude ranges from $i$ = 18 to 22 mag.
\replaced{Most of them}{Among 74 spectroscopic samples, 92\% of them} have $C$ values of --0.02 to 0.15, satisfying the GC selection criteria adopted in this study.
However, seven of them have C values of $0.15 < C < 0.25$, exceeding the GC selection criterion.
Therefore, these extended GCs will be missed in our GC selection.

	\begin{figure}[h]
\includegraphics[trim={0 1.5cm 0 1cm},clip,width=\columnwidth]{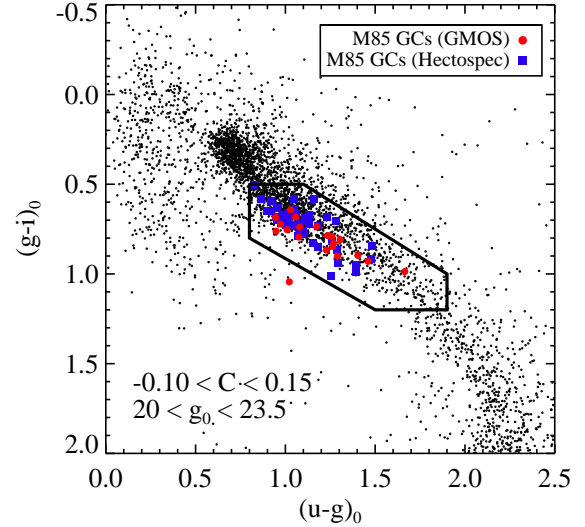}
\epsscale{1}
\caption{$(g-i)_0-(u-g)_0$ color-color diagram for the sources with $-0.10 < C < 0.15$ and 20 mag $< g_0 <$ 23.5 mag in the survey region.
The filled circles and squares are the same as in {\bf Figure \ref{fig:c_index}}.
The large polygon indicates the color criteria for the GC selection, as used in \citet{lim17}.
\label{fig:ccd}}
	\end{figure}
	
{\bf Figure \ref{fig:ccd}} shows the $(g-i)_0-(u-g)_0$ color-color diagram for the sources with $-0.10 < C < 0.15$ and 20.0 mag $< g_0 <$ 23.5 mag.
We used the foreground galactic extinction values for M85 derived by \citet{sf11}: $A_u = 0.128$, $A_g = 0.100$, and $A_i = 0.051$.
\citet{lim17} presented a guideline for the GC selection based on $ugi$ photometry using spectroscopically confirmed GCs in M87.
We adopted their color criteria, shown by a polygon in {\bf Figure \ref{fig:ccd}}, as follows:
\begin{displaymath}
0.5 < (g-i)_0 < 1.2,
\end{displaymath}
\begin{displaymath}
0.8 < (u-g)_0 < 1.9, 
\end{displaymath}
\begin{displaymath}
(g-i)_0 < 0.571 \times ((u-g)_0 - 0.8) + 0.8\rm{,~and}
\end{displaymath}
\begin{displaymath}
(g-i)_0 > 0.625 \times ((u-g)_0 - 1.1) + 0.5.
\end{displaymath}

We also plot the spectroscopically confirmed GCs in M85 (Ko et al. 2018; Ko, Y. et al. 2019, in preparation) in {\bf Figure \ref{fig:ccd}}. All except one are located inside the polygon, satisfying the GC selection criteria.
Note that there are few RGCs in the spectroscopically confirmed GCs 
of M85 compared to the red limit of the guideline based on M87 GCs.
\added{These color criteria are efficient in distinguishing GC candidates from background galaxies. The number of selected GC candidates remarkably decreases from 28,165 to 3,160.}
In addition to the color criteria, we set the magnitude criterion as 20.0 mag $< g_0 <$ 23.5 mag.
The point-like sources with $g_0 < 20.0$ mag are mainly contaminated by foreground stars.
We set the faint limit $g_0 = 23.5$ mag, where the $(g-i)_0$ color uncertainty is around 0.1 mag.

Applying all the criteria mentioned above, we selected 1318 GC candidates in the entire survey field.
A catalog of these GC candidates is presented in {\bf Table \ref{tab:gc.cat}}.

\begin{deluxetable*}{l l c c c c c c c c}
\tabletypesize{\footnotesize}
\tablecaption{A Photometric Catalog of the GC Candidates in M85 \label{tab:gc.cat}}
\setlength{\tabcolsep}{0.2in}
\tablewidth{\textwidth}
\tablehead{
\colhead{ID} & \colhead{$\alpha$ (J2000)} & \colhead{$\delta$ (J2000)} & \colhead{$g$} & \colhead{$u-g$} & \colhead{$g-i$} & \colhead{$C^a$} \\
\colhead{} & \colhead{(deg)} & \colhead{(deg)} & \colhead{(mag)} & \colhead{(mag)} & \colhead{(mag)} & \colhead{(mag)}
}
\startdata
GC0001 & 185.831070 & 18.427156 & 23.392 $\pm$ 0.032 & 1.03 $\pm$ 0.09 & 0.89 $\pm$ 0.05 & --0.046 \\
GC0002 & 185.832062 & 18.395197 & 22.541 $\pm$ 0.017 & 1.11 $\pm$ 0.05 & 0.61 $\pm$ 0.03 & --0.039 \\
GC0003 & 185.834488 & 18.306896 & 22.624 $\pm$ 0.014 & 0.96 $\pm$ 0.03 & 0.73 $\pm$ 0.02 & 0.087 \\
GC0004 & 185.835114 & 18.226645 & 21.112 $\pm$ 0.004 & 0.96 $\pm$ 0.01 & 0.69 $\pm$ 0.01 & --0.018\\
GC0005 & 185.835190 & 18.544601 & 22.120 $\pm$ 0.009 & 0.97 $\pm$ 0.02 & 0.83 $\pm$ 0.01 & 0.050 \\
\enddata
\tablecomments{{\bf Table \ref{tab:gc.cat}} is published in its entirety in the electronic edition.  
The five sample GCs are shown here as guidance for the table's form and content.}
\tablenotetext{a}{
The magnitude concentration index $C$ defined as the $i$-band magnitude differences between 4- and 8-pixel-diameter aperture photometry.}
\end{deluxetable*}

	\begin{figure}[hbt]
\includegraphics[trim={1cm 0.5cm 0 0},clip,width=\columnwidth]{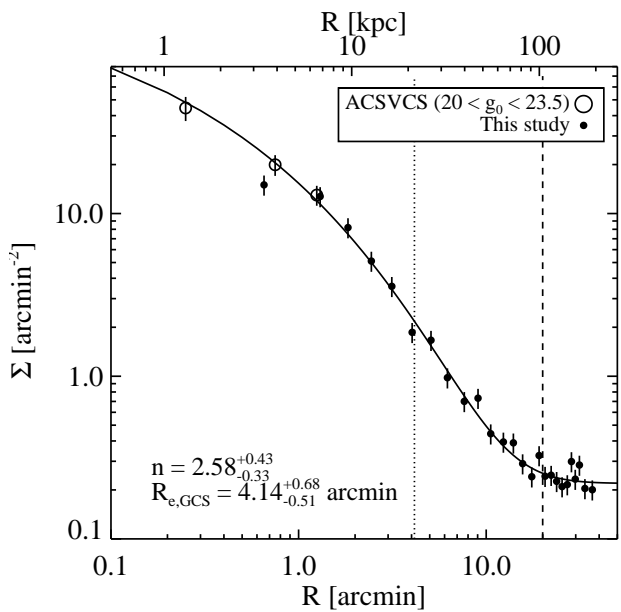}
\epsscale{1}
\caption{Radial number density profile of the GC candidates in the survey region.
The open and filled circles represent the GCs with 20 mag $< g_0 <$ 23.5 mag in the ACSVCS \citep{jor09} and the GC candidates detected in this study, respectively.
The solid line represents the best fit with a S{\'e}rsic function including the background level.
The vertical dotted and dashed lines represent the effective radius of the GC system and the boundary of the M85 GC system ($R = 20\arcmin$), respectively.
\label{fig:rd1}}
	\end{figure}

	\subsection{Radial Number Density Profiles of the GC Candidates}

In {\bf Figure \ref{fig:rd1}}, we derived the radial number density profile of the GC candidates in the survey region.
The radial surface density profile decreases as the galactocentric distance increases at $R < 20\arcmin$, and it becomes flat at $R > 20\arcmin$.
Thus, most of the GC candidates at $R < 20\arcmin$ are indeed \deleted{the} members of M85.

\replaced{We also plotted the radial number density profile of the GCs with $20 < g_0 < 23.5$ mag in the inner region at $R < 1\farcm5$ of M85 we derived from the data in the ACSVCS \citep{jor09}.}
{To supplement the radial density profile for the central region of M85, we used the ACSVCS photometry of GCs in M85 \citep{jor09}. From this photometry, we obtained the radial number density profile of the GCs with 20 mag $< g_0 <$ 23.5 mag at $R < 1\farcm5$, as displayed in {\bf Figure \ref{fig:rd1}}.}
The radial number density profile of the ACSVCS GCs \deleted{with $20 < g_0 < 23.5$ mag} at $1\arcmin < R < 1\farcm5$ shows good agreement with that in this study.
\replaced{The flattening of the radial profile of the CFHT GCs at $R < 1\arcmin$ is due to the incompleteness of our survey.}
{The radial profile of the CFHT GCs flattens at $R = 0\farcm65$ where the photometric completeness drops to 65\% (see {\bf Figure \ref{fig:completeness}}). After completeness correction, the number density becomes consistent with that derived from the ACSVCS data.}
We combined the ACSVCS data for the inner region at $R < 1\farcm5$ and the CFHT data for the outer region at $R > 1\farcm5$ to derive the radial number density profile of the entire region of M85.

Then, we fitted the radial number density profile of the GCs with a S{\'e}rsic function to measure the size of the GC system and the background level.
The S{\'e}rsic profile including the background level is defined as follows \citep{ser63,gd05}:
\begin{equation}
N(R) = N_{\rm e} ~ {\rm exp} \Bigg(-b_{\rm n}\Bigg[\Big(\frac{R}{R_{\rm e}}\Big)^{\frac{1}{n}}-1\Bigg]\Bigg)+{\rm bg},
\end{equation}
where $N_{\rm e}$ is the number density of the GCs at the effective radius $R_{\rm e}$, $n$ is the S{\'e}rsic index, $b_{\rm n}$ is the term related with $n$ ($b_{\rm n} = 1.9992n - 0.3271$), and bg is the background level.

The uncertainties of the parameters are estimated using the bootstrap procedure.
\replaced{We construct 1000 artificial data sets by randomly choosing 1318 GC candidates from among the actual data.}
{We constructed an artificial data set by randomly choosing 1318 GC candidates from the actual data allowing replacement, and we repeated it 1000 times to create 1000 artificial data sets.}
We fitted the radial number density profile based on the artificial data sets with the S{\'e}rsic function, and we estimated 16th and 84th percentiles for the fitted parameters.
The differences between these values and the parameters derived using the actual data are defined as the uncertainties of the parameters.
We derived the S{\'e}rsic index to be $n = 2.58^{+0.43}_{-0.33}$, the effective radius of the GC system to be $R_{\rm e,GCS} = 4.14^{+0.68}_{-0.51}$ arcmin, and the background level to be bg = $0.22^{+0.01}_{-0.02}$ arcmin$^{-2}$.
The S{\'e}rsic fit parameters are summarized in {\bf Table \ref{tab:sersic}}.
The GC candidates at $R < 20\arcmin$, corresponding to $R < 5R_{\rm e,GCS}$, are considered to be mostly genuine GCs, \added{of which the number is 810}. We used these sources for the following analysis.
\deleted{The number of the GC candidates within $R = 20\arcmin$ is 810.
After background subtraction (with the background level 0.22 arcmin$^{-2}$), it is expected that 533 GCs within $R = 20\arcmin$ belong to M85.}

\replaced{We estimate the total number of the GCs in M85 based on the S{\'e}rsic function derived above and the GC luminosity function (GCLF) of M85.
The total number of the GCs with $20 < g_0 < 23.5$ mag in M85 is estimated to be about 617$^{+47}_{-59}$ by integrating the background-subtracted S{\'e}rsic profile.
To estimate the number of the bright GCs with $g_0 < 20$ mag and faint GCs with $g_0 > 23.5$ mag, we adopt the GCLF of M85 derived by the ACSVCS.
\citet{jor07} presented the GCLFs of Virgo early-type galaxies using the ACSVCS data.
The GCLF of M85 follows a Gaussian distribution with a peak of $g_0 = 25.120$ and a width of 1.7.
This Gaussian width is much broader than that of the GCLFs of other early-type galaxies.
This is because the GCLF of M85 shows a faint excess due to a number of diffuse star clusters \citep{pen06b}.
We estimate the number fraction of the GCs with $20 < g_0 < 23.5$ mag to the total number using the Gaussian fit results.
As a result, the total number of the M85 GCs with the entire magnitude range and radial range is estimated to be about $N(total) \sim 3600^{+276}_{-341}$.
Adopting the total $V$-band luminosity of M85, $M_V = -22.34$ mag \citep{pen08}, we derive the specific frequency of the M85 GC system, $S_{\rm N} = 4.17^{+0.32}_{-0.40}$.
The uncertainties of the total number and specific frequency of the M85 GCs are estimated using the bootstrap procedure based on 1000 artificial data, as described above.

The mean specific frequency of Virgo early-type galaxies with $-24 < M_V < -22$ mag in the ACSVCS sample is 4.0 \citep{pen08}, which is comparable with the specific frequency of M85 GCs derived in this study.
\citet{pen08} constructed the radial number density profile of the M85 GCs using the ACSVCS and WFPC2 images taken near M85, and derived the specific frequency of the M85 GCs as $S_{\rm N}$ = 1.29 $\pm$ 0.21.
This value is much smaller than one derived in this study.
This large difference may be due to the uncertainty in the HST data that covered \added{a} much smaller field than the CFHT field in this study.
Our CFHT images covered $R \sim 30\arcmin$, while the HST images covered only $R \sim 7-8\arcmin$.}
{
\begin{deluxetable}{l c c c c c c c}
\tabletypesize{\footnotesize}
\tablecaption{S{\'e}rsic Fit Parameters for the Radial Number Density Profiles of the GC Systems \label{tab:sersic}}
\setlength{\tabcolsep}{0.02in}
\tablewidth{\columnwidth}
\tablehead{
\colhead{Sample} & \colhead{$N_{\rm e}$} & \colhead{$n$} & \colhead{$R_{\rm e}$} & \colhead{$R_{\rm e}$} & \colhead{bg} \\
\colhead{} & \colhead{(arcmin$^{-2}$)} & \colhead{} & \colhead{(arcmin)} & \colhead{(kpc)} & \colhead{(arcmin$^{-2}$)}
}
\startdata
All GCs & 1.96$^{+0.51}_{-0.51}$ & 2.58$^{+0.43}_{-0.33}$ & 4.14$^{+0.68}_{-0.51}$ & 21.6$^{+3.5}_{-2.7}$ & 0.22$^{+0.01}_{-0.02}$ \\
BGCs & 0.85$^{+0.31}_{-0.22}$ & 2.28$^{+0.41}_{-0.36}$ & 5.75$^{+0.91}_{-0.96}$ & 30.0$^{+4.7}_{-5.0}$ & 0.14$^{+0.01}_{-0.02}$ \\
RGCs & 1.98$^{+0.02}_{-0.80}$ & 1.80$^{+0.58}_{-0.54}$ & 2.02$^{+0.47}_{-0.06}$ & 10.5$^{+2.4}_{-0.3}$ & 0.09$^{+0.01}_{-0.01}$ \\
\enddata

\end{deluxetable}
	
	\subsection{Luminosity Function and Total Number of GCs}
 
{\bf Figure \ref{fig:lf}(a) and (b)} show the GCLFs for the galaxy region and the background region.
We corrected the GCLFs using the completeness test results derived in Section 2.3 and derived the background-subtracted GCLF considering the ratio of their areas (see {\bf Figure \ref{fig:lf}(c)}).
We fitted the background-subtracted GCLF with a Gaussian function, using only the magnitude range of $g_0 < 24.32$ mag where the completeness is higher than 50\%.
The turnover magnitude and Gaussian width were derived to be $g_0 = 23.48 \pm 0.07$ mag and $\sigma = 1.05 \pm 0.05$, respectively.

	\begin{figure}[t]
\includegraphics[width=\columnwidth]{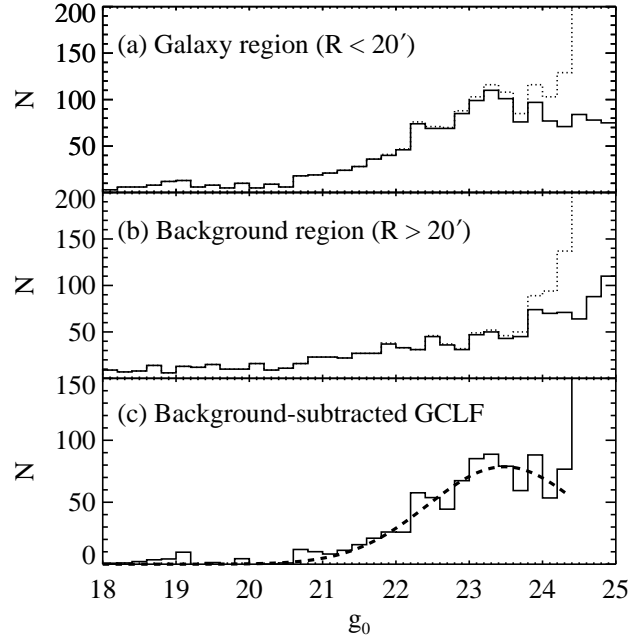}
\epsscale{1}
\caption{\added{GCLFs for (a) the galaxy region ($R < 20\arcmin$) and (b) the background region ($R > 20\arcmin$).
The solid and dotted histograms represent the GCLF before and after completeness correction, respectively.
(c) Background-subtracted GCLF. The dashed line shows a Gaussian fitting for the magnitude range of $g_0 < 24.32$ mag.
}
\label{fig:lf}}
	\end{figure}
	
\citet{jor07} presented the GCLFs of Virgo early-type galaxies using the ACSVCS data.
They found that the GCLF of M85 follows a Gaussian distribution with a peak of $g_0 = 25.120$ mag and a width of $\sigma = 1.7$.
The turnover magnitude and Gaussian width derived in this study are much brighter and narrower, respectively, than those from \citet{jor07}.
This difference is due to the faint excess at $g_0 > 24.5$ mag in the GCLF of the ACSVCS, which is not detected in this study.
M85 is known to host a number of diffuse star clusters 
that have low surface brightness of $\mu_g \geq 20$ mag arcsec$^{-2}$ \citep{pen06b}. 
\citet{jor07} pointed out that the faint end of the ACSVCS GCLF might be contaminated by faint diffuse star clusters.
However, we only considered typical GCs brighter than $g_0 = 24.32$ 
mag when constructing the GCLF in this study.

We estimated the total number of GCs in M85 based on their radial number density profile and the GCLF derived in this study.
The radial number density profile of the GCs with 20 mag $< g_0 <$ 23.5 mag was fitted to the S{\'e}rsic function including the background level (see Section 3.2 and {\bf Figure \ref{fig:rd1}}).
We integrated the background-level-subtracted S{\'e}rsic function from $R = 0$ up to $R = 20\arcmin$, estimating the number of GCs with 20 mag $< g_0 <$ 23.5 mag to be 617$^{+47}_{-59}$.
We calculated the number fractions of the bright GCs with $g_0 < 20$ mag and the faint GCs with $g_0 > 23.5$ mag using the Gaussian fit result of the GCLF, which are 0.05\% and 49\%, respectively.
Adding up the number of GCs in the entire magnitude range, the total number of GCs was estimated to be $N$(total) 
 $=1216^{+82}_{-50}$.
Adopting the total $V$-band luminosity of M85, $M_V = -22.34$ mag \citep{pen08}, we derived the specific frequency of the M85 GC system to be $S_{\rm N} = 1.41^{+0.10}_{-0.06}$.
The uncertainties of the total number and specific frequency of M85 GCs were estimated using the bootstrap procedure, as described above.

The specific frequency of the M85 GCs presented by \citet{pen08} is consistent with the one derived in this study within uncertainties. However, it is hard to directly compare these two values because we did not include low surface brightness star clusters in this study, and the ACSVCS did not cover the entire radial range of M85.
}

	\begin{figure}[hbt]
\includegraphics[trim={0 0.8cm 0 0},clip,width=\columnwidth]{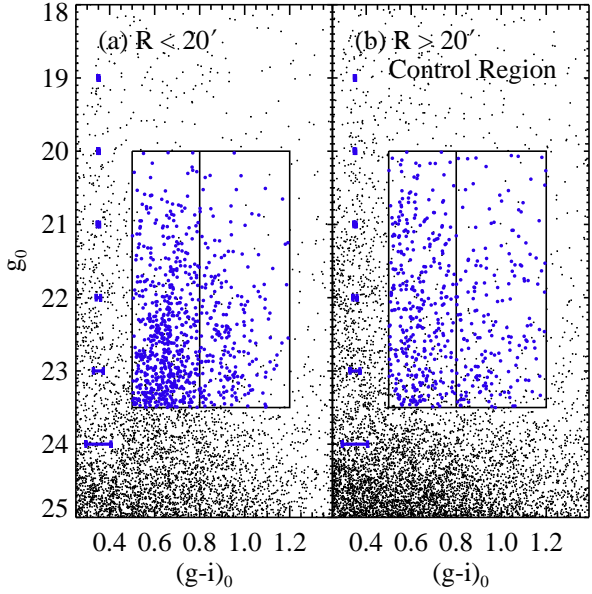}
\epsscale{1}
\caption{$g_0-(g-i)_0$ CMDs for the sources with $-0.10 < C < 0.15$ (gray dots) and GC candidates (blue filled circles) in (a) the M85 region ($R < 20\arcmin$) and (b) the control region ($R > 20\arcmin$).
The large boxes are the color and magnitude criteria for the GC selection, and the vertical line at $(g-i)_0 = 0.8$ indicates the criterion for selecting BGCs and RGCs.
The error bars on the left side in the CMDs represent the mean color errors of the point sources in each magnitude bin.
\label{fig:cmd1}}
	\end{figure}

	\begin{figure}[hbt]
\includegraphics[trim={0 0.8cm 0 0},clip,width=\columnwidth]{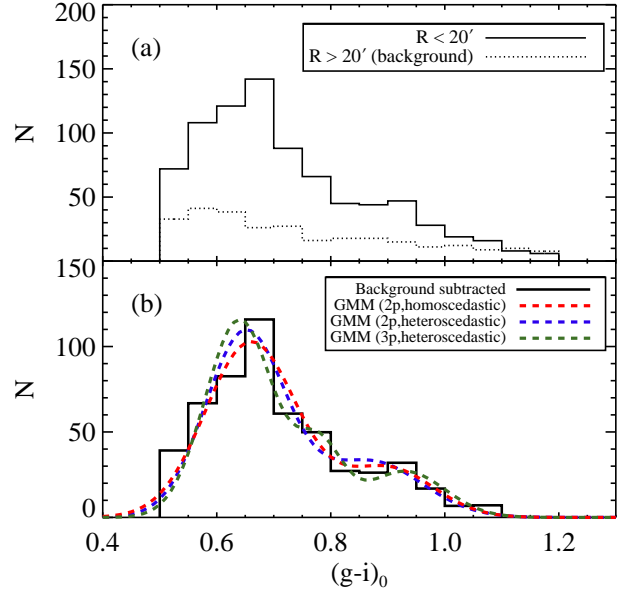}
\epsscale{1}
\caption{(a) $(g-i)_0$ color distribution of the GC candidates 
at $R < 20\arcmin$ (solid line) and $R > 20\arcmin$ (dotted line).
(b) Background-subtracted $(g-i)_0$ color distribution of the GCs 
at $R < 20\arcmin$ (solid line).
The red and blue dashed lines show the Gaussian fits from GMM in the bimodal case with the same variances and different variances, respectively.
The green dashed line shows the Gaussian fits from GMM in the trimodal case.
\label{fig:cdist}}
	\end{figure}
	
	\subsection{Color Distribution of the GC Candidates}

{\bf Figure \ref{fig:cmd1}(a) and (b)} show $g_0 - (g-i)_0$ color-magnitude diagrams (CMDs) for the sources with $-0.10 < C < 0.15$ in the galaxy region at $R < 20\arcmin$ and in the control region 
at $R > 20\arcmin$.
We marked the magnitude and color boundaries for GC selection by large boxes and plotted the GC candidates by blue symbols.
The M85 GC candidates show a distinct sequence on the CMD, which is not seen in the control region. 
In particular, a strong vertical sequence appears at $(g-i)_0 \sim 0.65$.
The GC candidates in the control region 
are mostly foreground stars.
The GC candidates with $(g-i)_0 < 0.8$ are mainly contaminated by the Sgr main-sequence, subgiant, and RGB stars, and the GC candidates with $(g-i)_0 > 0.8$ are mainly contaminated by Milky Way disk stars \citep{dur14}.

{\bf Figure \ref{fig:cdist}(a)} shows the $(g-i)_0$ color distribution of the GC candidates.
The GC candidates in the control region 
($R > 20\arcmin$) have a more uniform color distribution with a bluer peak than those in M85.
We subtracted the foreground and background contamination from the color distribution of GCs at $R < 20\arcmin$, considering the ratio of their areas.
The background-subtracted color histogram of the GCs at $R < 20\arcmin$ ranges from $(g-i)_0$ = 0.5 to 1.1, and shows two peaks, a strong peak at $(g-i)_0 = 0.675$ and a much weaker peak at $(g-i)_0 = 0.925$.

To decompose this background-subtracted color distribution of the GCs, we performed the Gaussian Mixture Modeling (GMM; Muratov \& Gnedin 2010), assuming bimodal color distributions with the same variances (homoscedastic case) and different variances (heteroscedastic case).
It is expected that there are about 277 contaminants among 810 GC candidates at $R < 20\arcmin$, according to the background level of bg = 0.22 arcmin$^{-2}$ from the radial number density profile fitting result (see Section 3.2).
We randomly picked 277 GC candidates in the background region.
Then, we subtracted 277 GC candidates in the galaxy region that have colors similar to those of the picked candidates in the background region.
We constructed 100 different data sets after subtracting the contaminants.
We ran GMM tests for these 100 data sets and measured the mean colors and Gaussian widths of the two components by adopting the mean values of the 100 GMM results.
We estimated the uncertainties of the mean colors and Gaussian widths corresponding to 68\% (1$\sigma$) confidence intervals. 
The GMM results are summarized in {\bf Table \ref{tab:gmm1}}.

\begin{deluxetable*}{c c c c c c c c}
\tabletypesize{\footnotesize}
\tablecaption{GMM Results for $(g-i)_0$ Color Distributions of the GCs in M85 \label{tab:gmm1}}
\setlength{\tabcolsep}{0.18in}
\tablewidth{\textwidth}
\tablehead{
\colhead{Sample} & \colhead{$N^{a}$} & \colhead{Mode} & \colhead{$\mu$} & \colhead{$\sigma$} & \colhead{$f$} & \colhead{$D$} & \colhead{$p$}
}
\startdata
\multirow{8}{*}{$R < 20\arcmin$} & \multirow{8}{*}{533} & Unimodal & 0.71$^{+0.01}_{-0.01}$ & 0.13$^{+<0.01}_{-0.01}$ & 1 & $\cdots$ & $\cdots$ \\
\cline{3-8}
& & Bimodal & 0.66$^{+0.01}_{-<0.01}$ & 0.08$^{+<0.01}_{-<0.01}$ & 0.78$^{+0.02}_{-0.02}$ & \multirow{2}{*}{3.10$^{+0.13}_{-0.12}$} & \multirow{2}{*}{4.27e-16} \\
& & (homoscedastic) & 0.90$^{+0.02}_{-0.02}$ & 0.08$^{+<0.01}_{-<0.01}$ & 0.22$^{+0.02}_{-0.02}$ & & \\
\cline{3-8}
& & Bimodal & 0.65$^{+0.01}_{-0.01}$ & 0.07$^{+0.01}_{-0.01}$ & 0.72$^{+0.06}_{-0.06}$ & \multirow{2}{*}{2.73$^{+0.51}_{-0.55}$} & \multirow{2}{*}{9.55e-16} \\
& & (heteroscedastic) & 0.87$^{+0.03}_{-0.03}$ & 0.09$^{+0.02}_{-0.02}$ & 0.28$^{+0.07}_{-0.06}$ & & \\
\cline{3-8}
& & \multirow{2}{*}{Trimodal} & 0.64$^{+0.01}_{-<0.01}$ & 0.06$^{+0.01}_{-<0.01}$ & 0.66$^{+0.09}_{-0.02}$ & \multirow{3}{*}{2.69$^{+0.21}_{-0.19}$} & \multirow{3}{*}{1.94e-14} \\
& & \multirow{2}{*}{(heteroscedastic)} & 0.78$^{+0.02}_{-0.01}$ & 0.04$^{+0.03}_{-0.02}$ & 0.15$^{+0.15}_{-0.10}$ & & \\
& & & 0.93$^{+0.01}_{-0.03}$ & 0.07$^{+0.02}_{-0.02}$ & 0.18$^{+0.06}_{-0.09}$ & & \\
\hline
\multirow{8}{*}{$R < 2\arcmin$} & \multirow{8}{*}{143} & Unimodal & 0.76$^{+<0.01}_{-<0.01}$ & 0.13$^{+<0.01}_{-<0.01}$ & 1 & $\cdots$ & $\cdots$ \\
\cline{3-8}
& & Bimodal & 0.68$^{+<0.01}_{-<0.01}$ & 0.09$^{+<0.01}_{-<0.01}$ & 0.64$^{+0.01}_{-0.01}$ & \multirow{2}{*}{2.37$^{+0.05}_{-0.05}$} & \multirow{2}{*}{2.18e-02} \\
& & (homoscedastic) & 0.89$^{+<0.01}_{-0.01}$ & 0.09$^{+<0.01}_{-<0.01}$ & 0.36$^{+0.01}_{-0.01}$ & & \\
\cline{3-8}
& & Bimodal & 0.68$^{+0.01}_{-0.01}$ & 0.08$^{+<0.01}_{-<0.01}$ & 0.59$^{+0.04}_{-0.04}$ & \multirow{2}{*}{2.28$^{+0.16}_{-0.11}$} & \multirow{2}{*}{1.01e-01} \\
& & (heteroscedastic) & 0.87$^{+0.01}_{-0.01}$ & 0.09$^{+0.01}_{-0.01}$ & 0.41$^{+0.04}_{-0.04}$ & & \\
\cline{3-8}
& & \multirow{2}{*}{Trimodal} & 0.67$^{+0.01}_{-0.01}$ & 0.08$^{+0.01}_{-<0.01}$ & 0.58$^{+0.08}_{-0.10}$ & \multirow{3}{*}{2.34$^{+0.18}_{-0.28}$} & \multirow{3}{*}{3.21e-01} \\
& & \multirow{2}{*}{(heteroscedastic)} & 0.85$^{+0.02}_{-0.02}$ & 0.07$^{+0.02}_{-0.01}$ & 0.35$^{+0.12}_{-0.09}$ & & \\
& & & 1.00$^{+0.01}_{-<0.01}$ & 0.04$^{+0.01}_{-0.02}$ & 0.07$^{+0.02}_{-0.03}$ & & \\
\hline
\multirow{8}{*}{$2\arcmin < R < 4\arcmin$} & \multirow{8}{*}{125} & Unimodal & 0.72$^{+<0.01}_{-<0.01}$ & 0.12$^{+<0.01}_{-<0.01}$ & 1 & $\cdots$ & $\cdots$ \\
\cline{3-8}
& & Bimodal & 0.66$^{+<0.01}_{-<0.01}$ & 0.07$^{+<0.01}_{-<0.01}$ & 0.73$^{+0.02}_{-0.01}$ & \multirow{2}{*}{3.13$^{+0.11}_{-0.10}$} & \multirow{2}{*}{1.25e-05} \\
& & (homoscedastic) & 0.89$^{+0.01}_{-0.01}$ & 0.07$^{+<0.01}_{-<0.01}$ & 0.27$^{+0.01}_{-0.02}$ & & \\
\cline{3-8}
& & Bimodal & 0.66$^{+0.03}_{-0.02}$ & 0.07$^{+0.02}_{-0.01}$ & 0.68$^{+0.18}_{-0.14}$ & \multirow{2}{*}{2.95$^{+0.92}_{-0.72}$} & \multirow{2}{*}{8.20e-05} \\
& & (heteroscedastic) & 0.88$^{+0.07}_{-0.05}$ & 0.07$^{+0.03}_{-0.04}$ & 0.32$^{+0.14}_{-0.18}$ & & \\
\cline{3-8}
& & \multirow{2}{*}{Trimodal} & 0.63$^{+<0.01}_{-<0.01}$ & 0.06$^{+<0.01}_{-<0.01}$ & 0.58$^{+0.03}_{-0.02}$ & \multirow{3}{*}{2.79$^{+0.14}_{-0.15}$} & \multirow{3}{*}{1.62e-04} \\
& & \multirow{2}{*}{(heteroscedastic)} & 0.79$^{+0.01}_{-0.01}$ & 0.05$^{+0.01}_{-0.01}$ & 0.27$^{+0.03}_{-0.03}$ & & \\
& & & 0.94$^{+<0.01}_{-0.01}$ & 0.04$^{+0.02}_{-0.01}$ & 0.15$^{+0.02}_{-0.02}$ & & \\
\hline
\multirow{8}{*}{$4\arcmin < R < 6\arcmin$} & \multirow{8}{*}{78} & Unimodal & 0.69$^{+0.01}_{-0.02}$ & 0.12$^{+0.01}_{-0.01}$ & 1 & $\cdots$ & $\cdots$ \\
\cline{3-8}
& & Bimodal & 0.64$^{+0.01}_{-0.01}$ & 0.07$^{+<0.01}_{-<0.01}$ & 0.78$^{+0.02}_{-0.02}$ & \multirow{2}{*}{3.43$^{+0.24}_{-0.24}$} & \multirow{2}{*}{6.21e-04} \\
& & (homoscedastic) & 0.87$^{+0.02}_{-0.01}$ & 0.07$^{+<0.01}_{-<0.01}$ & 0.22$^{+0.02}_{-0.02}$ & & \\
\cline{3-8}
& & Bimodal & 0.65$^{+0.01}_{-0.01}$ & 0.08$^{+0.01}_{-<0.01}$ & 0.83$^{+0.04}_{-0.02}$ & \multirow{2}{*}{3.93$^{+0.30}_{-0.21}$} & \multirow{2}{*}{2.37e-03} \\
& & (heteroscedastic) & 0.90$^{+0.02}_{-0.01}$ & 0.04$^{+0.01}_{-0.01}$ & 0.17$^{+0.02}_{-0.04}$ & & \\
\cline{3-8}
& & \multirow{2}{*}{Trimodal} & 0.60$^{+0.04}_{-0.04}$ & 0.05$^{+0.03}_{-0.02}$ & 0.48$^{+0.33}_{-0.25}$ & \multirow{3}{*}{2.57$^{+0.93}_{-0.66}$} & \multirow{3}{*}{4.25e-03} \\
& & \multirow{2}{*}{(heteroscedastic)} & 0.75$^{+0.09}_{-0.07}$ & 0.06$^{+0.02}_{-0.03}$ & 0.40$^{+0.25}_{-0.31}$ & & \\
& & & 0.92$^{+0.01}_{-0.01}$ & 0.03$^{+0.02}_{-0.01}$ & 0.12$^{+0.06}_{-0.04}$ & & \\
\enddata
\tablecomments{$^{a}$ The background-subtracted number of GCs.}
\end{deluxetable*}

In addition, we tried to decompose the color distribution of the GCs into three components with different variances to check the presence of intermediate-color GCs in M85.
The $p$ value indicates the probability for the unimodal distribution, and the $D$ value is the peak separation relative to the Gaussian widths.
If the $p$ value is smaller than 0.0001 and the $D$ value is greater than 2, it means that the input distribution is not unimodal and has a clear peak separation.

According to the derived $p$ and $D$ values for our three GMM results, we conclude that the color distribution of the M85 GCs is not unimodal.
The fitting results of two bimodal cases are consistent within uncertainties.
Therefore, we adopted the GMM results with the homoscedastic assumption for the bimodal case in the following analysis.
The mean colors of the two subpopulations are $(g-i)_0$ = 0.66 and 0.90, and the Gaussian widths are 0.08.
For the trimodal case, the mean colors of the three subpopulations are $(g-i)_0$ = 0.64, 0.78, and 0.93.
According to GMM results, the intermediate-color GCs constitute 15$^{+15}_{-10}$\% of the entire GCs, but the uncertainties are too large to confirm the presence of such a subpopulation.

	\begin{figure}[t]
\includegraphics[trim={1.5cm 0.5cm 0 1cm},clip,width=\columnwidth]{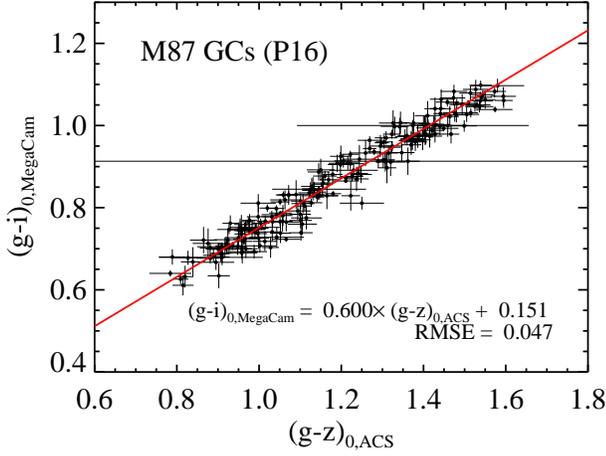}
\epsscale{1}
\caption{Color relation between $(g-z)_0$ \citep{jor09} and $(g-i)_0$ \citep{pow16} for the GCs in M87.
The solid line represents the least-squares fit between the two color indices.
\label{fig:comp_color}}
	\end{figure}
	
We compared the mean colors of GC subpopulations in M85 with those in other Virgo early-type galaxies.
\citet{pen06a} presented the relation between the mean colors of GC subpopulations and the absolute $B$-band magnitude of their host galaxy based on ACSVCS data.
According to this relation, the mean $(g-z)_{0, {\rm ACS}}$ colors of the BGCs and RGCs in the galaxies with the luminosity of M85 ($M_B = -21.09$; Binggeli et al. 1985) are derived to be 0.942 $\pm$ 0.109 and 1.386 $\pm$ 0.027.

To transform this $(g-z)_{0, {\rm ACS}}$ color in the ACS magnitude system to $(g-i)_{0, {\rm MegaCam}}$ color in the MegaCam magnitude system, we used the photometric data of the GCs in M87 presented by \citet{pow16}.
They selected a robust sample of GCs in the core region of the Virgo Cluster based on photometric data from NGVS and its near-infrared counterpart NGVS-IR \citep{mun14}.
Their photometry was tied to the CFHT/MegaCam AB magnitude system.
There are 215 GCs common to the GC catalogs of \citet{pow16} and the ACSVCS \citep{jor09}.
{\bf Figure \ref{fig:comp_color}} displays the color relation between $(g-i)_{0, {\rm MegaCam}}$ and $(g-z)_{0, {\rm ACS}}$ for those GCs in M87, and they show a tight correlation between the two colors.
We derived a transformation relation between $(g-i)_{0, {\rm MegaCam}}$ and $(g-z)_{0, {\rm ACS}}$,
\begin{displaymath}
(g-i)_{0,{\rm MegaCam}} = 0.600 (\pm 0.009) \times (g-z)_{0,{\rm ACS}} + 0.151 (\pm 0.010).
\end{displaymath}
The rms scatter in this relation is 0.047 mag.
According to this relation, the mean $(g-i)_0$ colors, converted from $(g-z)_{0, {\rm ACS}}$ colors, of BGCs and RGCs in the other Virgo galaxies 
that are as luminous as M85 
were derived to be $(g-i)_0$ = 0.716$^{+0.085}_{-0.083}$ and 0.983 $\pm$ 0.039, respectively.
The mean color of the RGCs in M85 ($(g-i)_0 = 0.90$) is bluer than those for the other Virgo early-type galaxies 
with a 2$\sigma$ or larger difference, while the mean color of the BGCs in M85 ($(g-i)_0 = 0.66$) is consistent within uncertainties. 
We discuss this result in Section 4.3.

	\begin{figure}[t]
\includegraphics[trim={0 0.5cm 0 0},clip,width=\columnwidth]{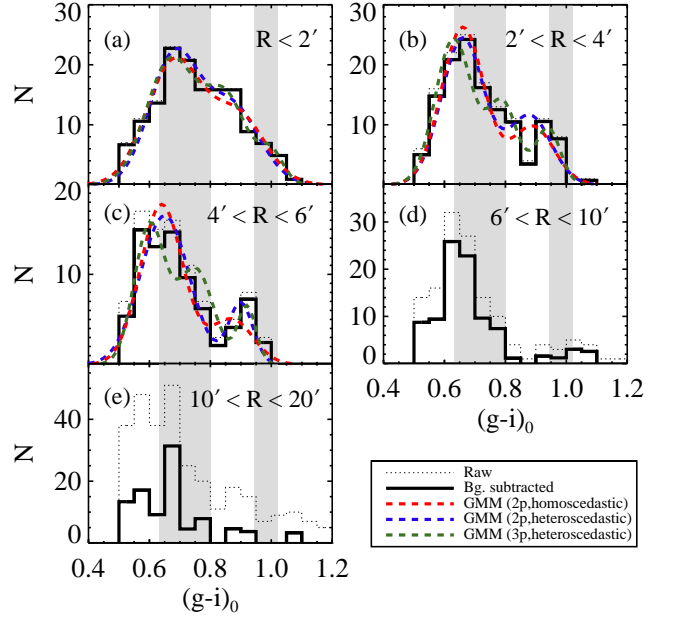}
\epsscale{1}
\caption{$(g-i)_0$ color distribution of the GCs in the different radial bins: (a) $R < 2\arcmin$, (b) $2\arcmin < R < 4\arcmin$, (c) $4\arcmin < R < 6\arcmin$, (d) $6\arcmin < R < 10\arcmin$, and (e) $10\arcmin < R < 20\arcmin$.
The dotted and solid histograms represent the raw and background-subtracted color distribution, respectively.
The dashed lines overplotted on the color histogram in panels (a), (b), and (c) are the same as in Figure \ref{fig:cdist}.
The gray shaded regions in all panels indicate the mean colors of BGCs and RGCs, $(g-i)_0$ = 0.716$^{+0.085}_{-0.083}$ and 0.983 $\pm$ 0.039, respectively, in Virgo early-type galaxies with luminosities similar to that of M85.
\label{fig:rbin_cdist}}
	\end{figure}
	
	\subsection{Radial Variation of GC Colors}	

{\bf Figure \ref{fig:rbin_cdist}} shows the $(g-i)_0$ color distributions of the GC candidates in different radial bins, $R < 2\arcmin$, $2\arcmin < R < 4\arcmin$, $4\arcmin < R < 6\arcmin$, $6\arcmin < R < 10\arcmin$, and $10\arcmin < R < 20\arcmin$.
We removed the background contamination from each color histogram.
The background-subtracted color distribution of the GC candidates 
at $R < 2\arcmin$ does not show 
  \replaced{clear two}{two clear} peaks, while those of the GC candidates 
at $2\arcmin < R < 4\arcmin$ and $4\arcmin < R < 6\arcmin$ show \replaced{clear two}{two clear} peaks.
The GC candidates with $R > 6\arcmin$ are mostly blue with $(g-i)_0 < 0.8$.
We performed the GMM tests for the GC candidates 
at $R < 2\arcmin$, $2\arcmin < R < 4\arcmin$ and $4\arcmin < R < 6\arcmin$ with the same procedure as described in Section 3.4.
The GMM results for the color distribution of the GC candidates in these radial bins are summarized in {\bf Table \ref{tab:gmm1}}.

	\begin{figure}[t]
\includegraphics[trim={0.5cm 0.5cm 0 0},clip,width=\columnwidth]{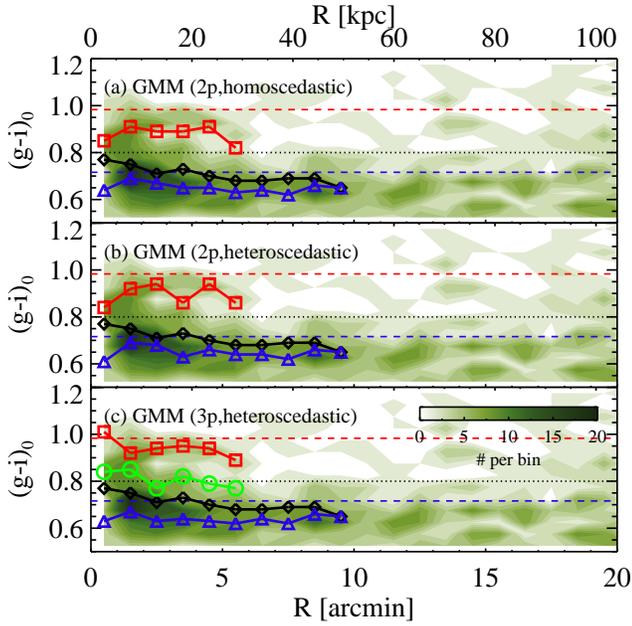}
\epsscale{1}
\caption{$(g-i)_0$ color vs. galactocentric distance for the GCs in M85. The contour represents the number density map of the GCs, and the color scale indicates the number of the GCs in radial and color bins.
The mean colors of all GCs (black diamonds), BGCs (blue triangles), and RGCs (red squares) as a function of galactocentric distance are overplotted.
In panel (c), the mean colors of intermediate-color GCs are also marked by green circles.
The two dashed horizontal lines at $(g-i)_0$ = 0.716 and 0.983 show the mean colors of BGCs and RGCs in Virgo early-type galaxies with  luminosities similar to that of M85.
The dotted horizontal line at $(g-i)_0 = 0.8$ indicates the criterion for separating BGCs and RGCs.
\label{fig:rcolor}}
	\end{figure}
	
We examined the possibility of a multimodal color distribution of GC candidates within $R = 10\arcmin$ in each radial bin.
The GC candidates 
at $R < 2\arcmin$ do not clearly show a 
color multimodality ($p = 0.02$ for the bimodal homoscedastic case, 0.10 for the bimodal heteroscedastic case, 0.32 for the trimodal case).
On the other hand, the GC candidates 
at $2\arcmin < R < 4\arcmin$ and $4\arcmin < R < 6\arcmin$ show clearly a 
multimodal color distribution ($p < 0.001$ and $D > 2$).
It is difficult to tell whether the color distribution of the GC candidates 
at $2\arcmin < R < 4\arcmin$ is bimodal or trimodal.
However, the GC candidates 
at $4\arcmin < R < 6\arcmin$ have a bimodal color distribution because the uncertainty of the number fraction of intermediate-color GCs for the trimodal case is too large to confirm their presence.
From these results, we 
conclude that the intermediate-color GCs exist within $R = 4\arcmin$ (=21 kpc).
	
We investigated the radial gradients of the mean $(g-i)_0$ colors of the BGC and RGC candidates in M85 (see {\bf Figure \ref{fig:rcolor}}).
Most GC candidates have $(g-i)_0$ colors bluer than 1.0.
We found two distinct sequences at $(g-i)_0$ = 0.65 and 0.95, corresponding to the BGC and RGC populations.
The BGC candidates are distributed over a 
wide area, but the RGC candidates are mostly located in the inner region at $R < 6\arcmin$.

We decomposed the color distribution of the GC candidates in each radial bin with a 1$\arcmin$ interval into two or three components using GMM tests as described in Section 3.4.
The radial color variations of the GC candidates are shown in {\bf Figure \ref{fig:rcolor}}.
The mean color of all GC candidates at the center of M85 is $(g-i)_0 = 0.77$ and becomes 0.12 mag bluer in the outer region at $R = 10\arcmin$.
From all GMM results, we did not find any significant color gradients for both BGC and RGC candidates.

	\begin{figure}[t]
\includegraphics[trim={0.5cm 0.8cm 0.5cm 1cm},clip,width=\columnwidth]{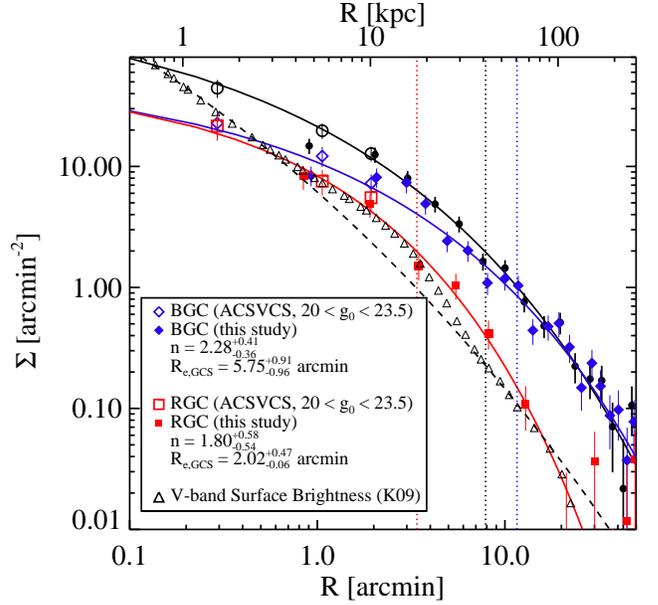}
\epsscale{1}
\caption{Background-subtracted radial number density profiles of the BGC (diamonds) and RGC (squares) systems, compared with the entire GC system (circles).
The solid lines represent the S{\'e}rsic law fitting results for the GC systems.
The black, blue, and red vertical dotted lines represent the effective radii of the entire GC, BGC, and RGC systems, respectively.
The triangles and dashed line represent the $V$-band surface brightness profile shifted arbitrarily and the S{\'e}rsic law fitting result for the surface brightness profile \citep{kor09}, respectively.
\label{fig:rd2}}
	\end{figure}

	\begin{figure*}[t]
\epsscale{0.9}
\plotone{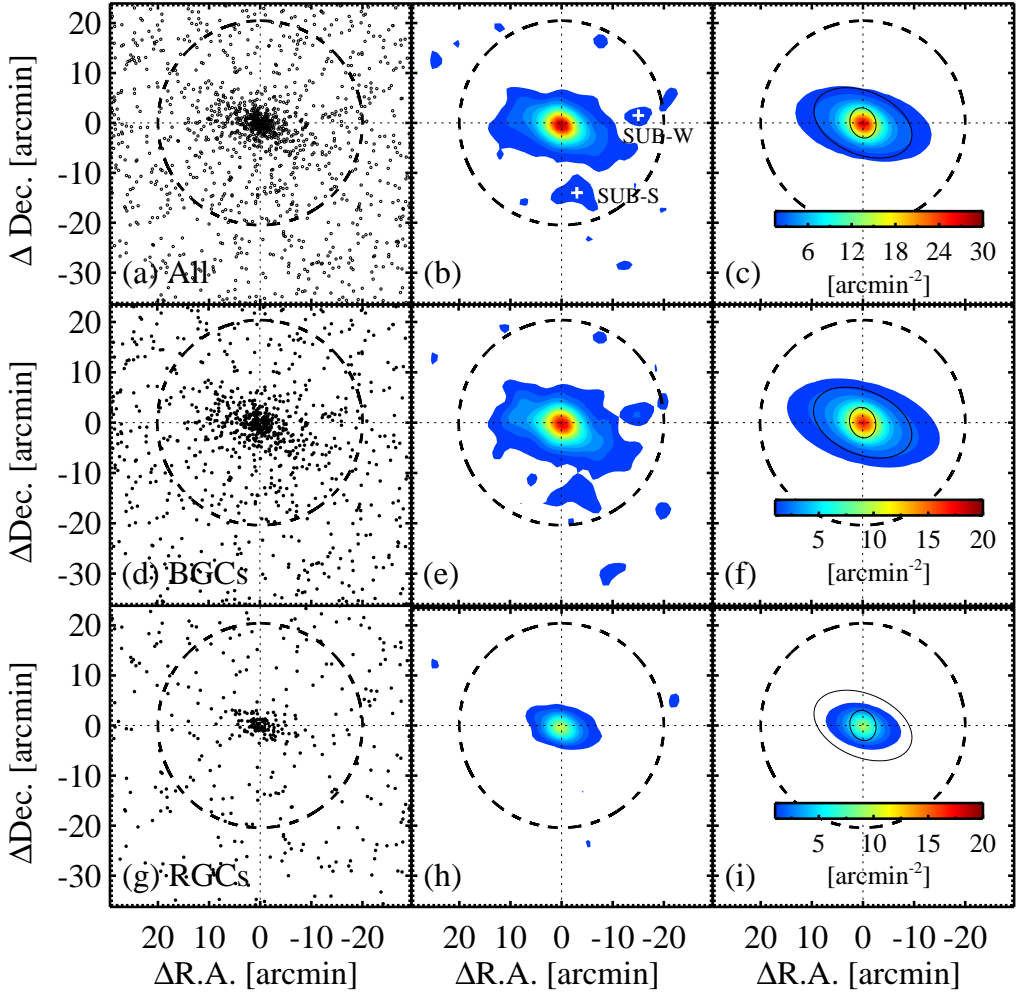}
\caption{(a) Spatial distributions of the GCs.
(b) Smoothed surface density map of the GCs, color-coded by the number density.
The crosses indicate the center positions of the substructures, SUB-S and SUB-W, detected in this study (see Section 3.7).
(c) Surface density map of the GCs modeled by the IRAF/ELLIPSE task.
The solid line ellipses indicate two guidelines for the stellar extent of M85 \citep{kor09}: One is with a major axis of 3 arcmin, a position angle of 16$^{\circ}$, and an ellipticity of 0.18, and the other is with a major axis of 10 arcmin, a position angle of 66$^{\circ}$, and an ellipticity of 0.38.
The dashed line circles represent the boundary of M85 GC system ($R=20\arcmin$).
(d)-(f) Same as (a)-(c) but for BGCs.
(g)-(i) Same as (a)-(c) but for RGCs.
\label{fig:spatial}}
	\end{figure*}

	\subsection{Spatial Distribution of the GC Candidates}

First, we investigated radial  distributions of all GC, BGC, and RGC candidates and compared them with the stellar light of M85.
To investigate any difference in the properties of the GCs depending on their colors, we divided the GC candidates into blue and red, with a color criterion $(g-i)_0 = 0.8$, which is the same as the one adopted in the NGVS \citep{dur14}.

We fitted the radial number density profiles of the BGC and RGC candidates in M85 with the same procedure as described in Section 3.2.
We subtracted the derived background level from the profiles and display the resulting profiles in {\bf Figure \ref{fig:rd2}}.
The S{\'e}rsic indices for the BGC and RGC systems were derived as $n = 2.28^{+0.41}_{-0.36}$ and $1.80^{+0.58}_{-0.54}$, respectively.
The slope of the radial number density profile of the RGCs in the outer region of M85 is much steeper than that of the BGCs.
The effective radii for the BGC and RGC systems were measured as $R_{\rm e,GCS} = 5.75^{+0.91}_{-0.96}$ arcmin ($\sim$30 kpc) and $2.02^{+0.47}_{-0.06}$ arcmin ($\sim$11 kpc), respectively.
Thus, the RGCs are more concentrated 
around the central region of M85 than the BGCs, which is commonly seen in the GC systems of early-type galaxies \citep{kis97,lkg98,str11,fpo12}.
The S{\'e}rsic fit parameters for each GC system are summarized in {\bf Table \ref{tab:sersic}}.
	
In addition, we compared the $V$-band surface brightness profile of M85 derived in \citet{kor09} with the radial number density profiles of the GCs.
\citet{kor09} presented the values of the S{\'e}rsic index and the effective radius for M85 derived from surface photometry: $n = 6.12^{+0.31}_{-0.27}$, $R_{\rm e,star} = 128.9^{+10.2}_{-8.8}$ arcsec 
from the major-axis fit, and $R_{\rm e,star} = 102.3\pm 6.3$ arcsec 
from 2D profile integration.
Noting the unusual features (e.g., an excess at $26\arcsec < R < 221\arcsec$) seen in the surface brightness profile of M85 (see their Figure 14), \citet{kor09} concluded that ``the galaxy is an elliptical -- a recent (damp?) merger remnant that has not fully settled into equilibrium."
The $V$-band surface brightness profile of the inner region ($R < 0\farcm3$) of M85 is steeper than the radial number density profile of the GCs, while the surface brightness profile of the outer region to $R = 20\arcmin$ is more correlated with the radial number density profile of the RGCs rather than that of the BGCs.
This supports that the RGCs were formed with the stars in the main body of M85, but the BGCs were brought by dissipationless mergers.
	
Next, we investigated two-dimensional spatial distributions of all GC, BGC, and RGC candidates 
 as shown in {\bf Figure \ref{fig:spatial}(a), (d), and (g)}. 
The GC candidates are concentrated around the center of M85.
{\bf Figure \ref{fig:spatial}(b), (e), and (h)} show the smoothed surface number density maps of the all GC, BGC, and RGC candidates.
The spatial distribution of the GC candidates is elongated roughly along the E and W directions.
The BGC system in M85 is extended out to $R \sim 20\arcmin$, while the RGC system is confined to a much smaller region at $R < 6\arcmin$.
We fitted the smoothed surface density maps using the IRAF/ELLIPSE task. Fixing the center position, we derived ellipticities and position angles of the isodensity contours.
The fitted models are shown in {\bf Figure \ref{fig:spatial}(c), (f), and (i)}.
The spatial distribution of the GC system in the outer region, $R \sim 10\arcmin$, is comparable to that of the galaxy light of M85 \citep{kor09}.

\begin{figure}[t]
\includegraphics[trim={1cm 0.8cm 0.5cm 0},clip,width=\columnwidth]{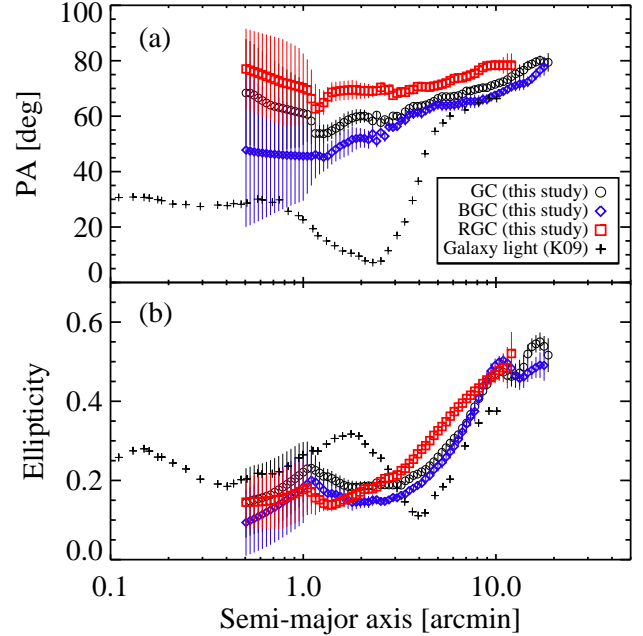}
\epsscale{1}
\caption{Comparison of (a) position angle and (b) ellipticity variations as a function of semi-major axis between the GC systems and the galaxy
 light \citep{kor09}.
The circles, diamonds, squares, and crosses represent the entire GC system, BGC system, RGC system, and galaxy light, respectively.
\label{fig:ellipse}}
	\end{figure}

{\bf Figure \ref{fig:ellipse}} shows the radial changes of the position angles and ellipticities of the isodensity contours of the GC systems of M85.
Note that few GCs are identified in the central region at $R < 0\farcm5$ in the CFHT/MegaCam images (see {\bf Figure \ref{fig:rd1}}), so we fitted the data for $R > 0\farcm5$.
We also plot the position angles and ellipticities of the galaxy light.
The position angles of the galaxy light isophotes dramatically increase from 5$^{\circ}$ to 65$^{\circ}$ in the semi-major axis range of 2$\arcmin$--10$\arcmin$, while those of the GC isodensity contours vary much more slowly from 60$^{\circ}$ to 80$^{\circ}$.
The ellipticities of the GC isodensity contours vary from 0.2 to 0.55 in the semi-major axis range of 1$\arcmin$--20$\arcmin$.
In the same region, the ellipticities of the galaxy light isophotes show some fluctuation between 0.1 and 0.4.
The position angles and ellipticities of the GC isodensity contours and the galaxy light isophotes change in a similar way at $R_{\rm maj} > 4\arcmin-5\arcmin$.
The isodensity contours of the BGCs and RGCs show similar features to those of all GCs.
This indicates that there are no significant differences in the shape between the two GC subsystems except for their radial extent.

	\begin{figure}[hbt]
\includegraphics[trim={1cm 0.5cm 0.5cm 0.5cm},clip,width=\columnwidth]{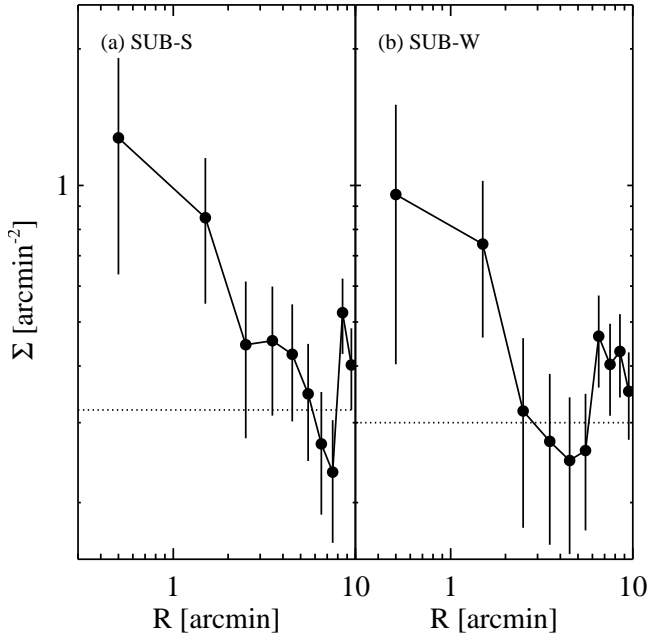}
\epsscale{1}
\caption{(a) Radial number density profile of the GCs in the SUB-S (filled circles).
The dotted horizontal line indicates that the number density of the M85 GCs at the SUB-S position.
(b) Same as panel (a), but for the SUB-W.
\label{fig:rd_sub}}
	\end{figure}	

	\begin{figure}[hbt]
\includegraphics[trim={0.5cm 0.5cm 0.5cm 0.5cm},clip,width=\columnwidth]{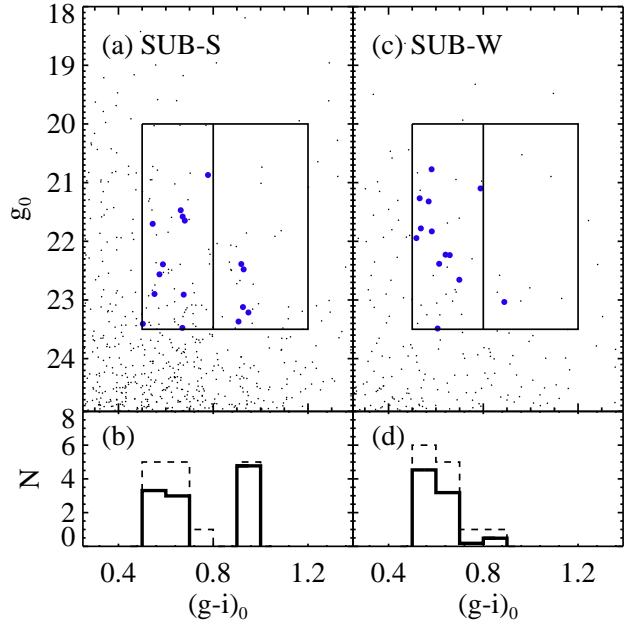}
\epsscale{1}
\caption{(a) $g_0-(g-i)_0$ CMD of the \replaced{point and compact}{point-like and slightly extended} sources in the SUB-S. The filled circles represent the GC candidates.
The large boxes and the vertical line at $(g-i)_0$ = 0.8 represent the same as in {\bf Figure \ref{fig:cmd1}}.
(b) $(g-i)_0$ color distribution of the GC candidates in the SUB-S (dashed line).
The solid histogram represents the background-subtracted $(g-i)_0$ color distribution of the GC candidates in the SUB-S.
Panels (c) and (d) are the same as panels (a) and (b), respectively, but for the SUB-W.
\label{fig:cmd_sub}}
	\end{figure}	

	\subsection{Substructures of the GC System}
	
In Figure \ref{fig:spatial}(b) and (e), we found two substructures at the southern and western regions from M85, called SUB-S and SUB-W, shown at ($\Delta$R.A.,$\Delta$decl.) = (--3$\arcmin$, --14$\arcmin$) and (--15$\arcmin$, 1$\farcm$5) and corresponding to ($\alpha$, $\delta$) = (186$\fdg$297882, 17$\fdg$958153) and (186$\fdg$087265, 18$\fdg$216486).
We checked the images to detect any diffuse light at the position of SUB-S and SUB-W, and we found little.
In addition, these substructures are not related to any other galaxies in the survey region.
The GC candidates in these two substructures are mostly blue, as clearly shown in {\bf Figure \ref{fig:spatial}(e)}.
{\bf Figure \ref{fig:rd_sub}} displays the radial number density profiles of the GC candidates from the center position of each substructure.
They show a significant central concentration, so that the GC candidates at the position of the substructures are mostly the members of each substructure.
The surface number densities of the M85 GCs at the SUB-S and SUB-W positions ($R \sim 14\farcm3$ [74.5 kpc] and $15\farcm1$ [78.7 kpc]) are 0.32 and 0.30 arcmin$^{-2}$, respectively, based on the S{\'e}rsic law fit result (see {\bf Figure \ref{fig:rd1}}).
We estimated the radial extent of both SUB-S and SUB-W to be $\sim 2\farcm5$ (13 kpc), where the surface number density of the GC candidates in each substructure reaches that of the M85 GCs within 1$\sigma$.
The number density excess at the outer part of each substructure (at $9'$ for SUB-S and $7'$ for SUB-W) is caused by the GCs that belong to M85.
The numbers of the GCs found in SUB-S and SUB-W are 16 and 13, respectively.

{\bf Figure \ref{fig:cmd_sub}(a) and (c)} show the $g_0-(g-i)_0$ CMDs of the GC candidates and the sources with $-0.10 < C < 0.15$ 
at a distance $<2\farcm5$ from the center of each of SUB-S and SUB-W.
The SUB-S and SUB-W are 14$\farcm$3 and 15$\farcm$1 away from M85, respectively.
We chose control regions at these radial ranges and subtracted the contribution of the M85 GCs in the control regions from the GC candidates in SUB-S and SUB-W.
{\bf Figure \ref{fig:cmd_sub}(b) and (d)} show the background-subtracted color distribution of the GCs in SUB-S and SUB-W.
The total numbers of the GCs in the SUB-S and SUB-W are 11 and 9, respectively, after background subtraction.

Assuming that the substructures and M85 are located at the same distance, and that the width of their GC luminosity function is about 1, we derived the total numbers of GCs in SUB-S and SUB-W, $N_{\rm total}$(GC) $\sim$ 209 and 171, respectively. 
If we use the relation between the total number of GCs and their host galaxy dynamical mass in \citet{hha13}, we obtain the progenitor masses for the substructures of $M_{\rm dyn} \sim 3-4 \times 10^{10} M_\odot$, indicating that they are candidates of dwarf galaxies.

\section{Discussion}

	\subsection{Elongated Spatial Distribution of the GC System}

According to the GC formation scenario suggested by \citet{cot98} and \citet{lee10}, the BGCs in massive early-type galaxies, especially in the outer region, mainly come from dwarf galaxies, and the RGCs formed along with the bulk of stars during dissipative mergers at the early epoch.
In this scenario, the BGCs show more extended spatial distributions than the RGCs \citep{kis97,lkg98,str11,fpo12}, and the spatial distribution of the RGCs shows a stronger correlation with that of the stellar light than that of the BGCs \citep{pl13,wan13}.

In the case of M85, the radial distribution of the BGCs is more extended than that of the RGCs (see {\bf Figure \ref{fig:rd2}}), but the BGCs in the outer region show a significant correlation with the stellar light, comparable to the RGCs (see {\bf Figure \ref{fig:ellipse}}).
Both the spatial distributions of the stellar light and the GC systems are elongated in the outermost region of M85.
Based on these results, we 
infer that the BGCs in M85 originated from dissipationless mergers, but after accretion the BGCs, RGCs, and stars in M85 underwent the same evolutionary process, resulting in the similar two-dimensional spatial distributions.

\citet{ko18} found that the M85 GC system \replaced{in the inner region of M85}{at $R < 3\arcmin$} strongly rotates with \deleted{a rotation amplitude of 235 km s$^{-1}$ and} a rotation axis angle of $\sim$161$^{\circ}$.
\replaced{The rotation axis angle of the GC system}{This value} is similar to the position angle of the photometric minor axis at $R > 10\arcmin$ ($\sim$170$^{\circ}$).
This agreement \deleted{between the rotation axis of the GC system and the photometric minor axis of stellar light} implies that the GC system at \added{$R > 10\arcmin$} may also have a strong rotation.
This will be investigated with kinematic data of GCs \replaced{in the wide field of M85}{at $R < 20\arcmin$} (Ko, Y. et al. 2019, in preparation). 
We suspect that an off-center major merging event might have occurred recently after the BGCs were accreted from low-mass galaxies.
		
	\subsection{Low RGC Fraction in the Central Region of M85}

The number fraction of the RGCs is expected to be related to the host galaxy mass according to current GC formation models \citep{az92,za93,fbg97,cot98,lee10}.
The BGCs were formed in progenitor spirals or low-mass dwarf galaxies at an early epoch and were accreted into massive galaxies via dissipationless mergers afterward, while the RGCs were formed in massive progenitors.
Consequently, the majority of the BGCs in massive early-type galaxies inhabit 
 the outskirts of the massive early-type galaxies, while the RGCs are mostly 
 located in the central region.
Massive early-type galaxies have undergone a number of wet mergers, resulting in the excess of the RGCs in the central region compared to the BGCs.
Therefore, the number fraction of the RGCs in the central region is expected to increase as a function of their host galaxy mass.

	\begin{figure}[t]
\includegraphics[trim={1cm 0.5cm 1cm 1cm},clip,width=\columnwidth]{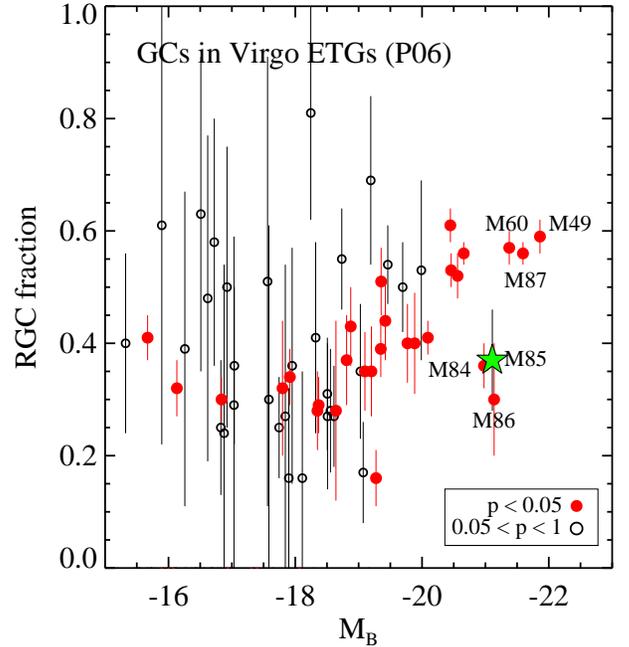}
\epsscale{1}
\caption{Number fraction of RGCs in Virgo early-type galaxies as a function of host galaxy luminosity \citep{pen06a}.
The filled and open circles represent the galaxies that show a bimodal color distribution of their GCs ($p < 0.05$) and the galaxies that do not show clear bimodality in their GC colors ($0.05 < p < 1$), respectively.
Note that M84, M85, and M86 show a 
much lower fraction of RGCs compared to other galaxies with similar luminosity.
\label{fig:fr_mb}}
	\end{figure}

\citet{pen06a} investigated the color distribution of the GCs in 100 Virgo early-type galaxies using ACSVCS data, and they presented a relation between the RGC fraction and host galaxy luminosity, as shown in {\bf Figure \ref{fig:fr_mb}}.
In the figure, we marked the galaxies that show a clear bimodality in GC color ($p < 0.05$) by red symbols.
The number fraction of the RGCs in the ACS field of these galaxies increases, in general, as a function of host galaxy luminosity.
However, there are three outliers at the high-luminosity end ($M_B \sim -21$): M84, M85, and M86.
These three galaxies have much lower RGC fractions, compared with the other Virgo early-type galaxies of similar luminosity (like M60).
\citet{pen06a} estimated the number fraction of M85 RGCs in the ACS field to be 37\%, although the color distribution of M85 GCs does not show a clear bimodality.
This value is similar to the number fraction of the RGCs at $R < 2\arcmin$ derived in this study, which is 36-41\% based on the bimodal GMM tests (see {\bf Table \ref{tab:gmm1}}).
This indicates that there are fewer RGCs or more BGCs in the central region of M85, compared to other Virgo early-type galaxies of similar luminosity.
This implies that the primary progenitors of M85, as well as M84 and M86, have been less massive or have been involved with fewer wet mergers than other massive galaxies of similar luminosity.

\added{	\subsection{Intermediate-color GCs}}

\deleted{In addition, the mean $(g-i)_0$ color of the RGCs is much bluer than that of early-type galaxies with similar luminosity (see \S3.3 and \S3.4).
To interpret these ``blue" RGCs or intermediate-color GCs, we consider the possibility of the presence of another population distinct from the old GCs.
For instance, \citet{lim17} investigated the GCs in NGC 474 and revealed that this galaxy lacks RGCs and have more intermediate-color GCs instead.
They interpreted that these intermediate-color GCs formed a few Gyr ago and eventually become the typical RGCs.}

\added{
We found two signatures for the presence of intermediate-color GCs in M85.
First, the color distribution of the GCs in the central region of M85 ($R < 4\arcmin$) is not simply bimodal. 
Second, the RGCs in the outer region of M85 have bluer colors than those in other early-type galaxies with similar luminosities to M85.

Not only M85 but several merger remnant galaxies are also known to host intermediate-color GCs.
These GCs are considered to be intermediate-age GCs that were formed during wet mergers a few gigayears ago.
NGC 474 is a shell galaxy that lacks RGCs and hosts more intermediate-color GCs instead \citep{lim17}.
These intermediate-color GCs are considered 
to have been formed a few gigayears ago, and they will eventually become typical RGCs.
NGC 3610 and NGC 4753 also host intermediate-color GCs \citep{cbg15,bc17}, which can be interpreted as 1--3 Gyr old GCs according to SSP models.
However, \citet{str03,sbf04} estimated the ages of 13 GCs in NGC 3610 from their spectra and found that two of the GCs are as young as $\sim$2 Gyr but they are as red as typical RGCs ($(V-I) \sim 1.2$).
There are no spectroscopic studies for NGC 4753 GCs in the literature.
In addition, NGC 1316 is a merger remnant galaxy in the outskirts of the Fornax Cluster, which is strikingly similar to M85 in several aspects.
The color distribution of the GCs in NGC 1316 indicates the presence of four different populations \citep{ric12,sff16}.
\citet{ses18} spectroscopically confirmed the presence of a dominant population of 2.1 Gyr old GCs among their intermediate-color GC samples.

On the other hand, there are some cases where the intermediate-color GCs are not simply explained as few-gigayear-old GCs.
For example, it has been uncertain whether NGC 4365 hosts intermediate-age GCs or not, while its GC color distribution is unimodal or trimodal \citep{kw01,lar01,bsf12}.
Some studies have suggested that NGC 4365 hosts intermediate-age GCs, based on near-infrared photometry \citep{puz02, hem07} and spectroscopy \citep{lar03}, while other studies have suggested that there are no signatures for the presence of intermediate-age GCs, based on near-infrared photometry \citep{chi11} and spectroscopy \citep{bro05, clk12}.
In addition, the GC color distribution of NGC 7507 shows a narrow peak at the intermediate color range \citep{cas13}.
However, these intermediate-color GCs are hardly interpreted as the intermediate-age GCs that were formed during wet mergers because this galaxy does not show 
any merging features.
}

	\begin{figure}[t]
\includegraphics[trim={1cm 0.5cm 1cm 1cm},clip,width=\columnwidth]{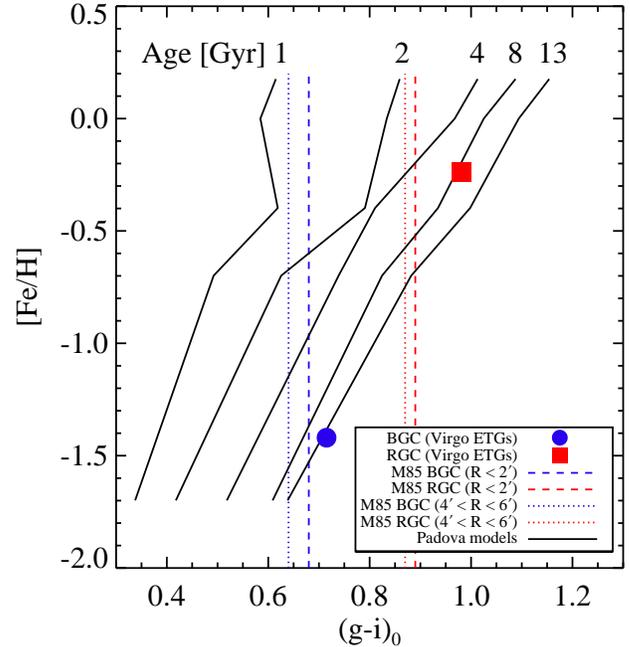}
\epsscale{1}
\caption{[Fe/H] vs. $(g-i)_0$ color for GCs.
The filled circle and square indicate the values for the BGCs and RGCs in the Virgo early-type galaxies with luminosities similar to that of M85, respectively, based on the relation between the mean GC colors and the absolute $B$ magnitude of the host galaxy \citep{pen06a}.
The mean $(g-i)_0$ colors of M85 BGCs and RGCs 
at $R < 2\arcmin$ (blue and red dashed lines) and $4\arcmin < R < 6\arcmin$ (blue and red dotted lines) are overplotted.
The black solid lines represent the relation between [Fe/H] and $(g-i)_0$ color given by the Padova simple stellar population models for ages of 1, 2, 4, 8, and 13 Gyr \citep{mar08,gir10}.
\label{fig:feh_color}}
	\end{figure}
	
Likewise, we investigated the ages and metallicities of the ``blue" RGCs in M85 based on the $(g-i)_0$ color and [Fe/H] relation.
In {\bf Figure \ref{fig:feh_color}} we plot, with the solid lines, the relation between [Fe/H] and $(g-i)_0$ color given by the Padova simple stellar population models for ages of 1, 2, 4, 8, and 13 Gyr \citep{mar08,gir10}.
\citet{pen06a} presented the relation between the mean [Fe/H] values of the GC subpopulations and host galaxy luminosity based on the ACSVCS sample that covered the central regions of galaxies.
According to this relation, the mean metallicities of the BGCs and RGCs in the other early-type galaxies that are as luminous as M85 are expected to be [Fe/H] = --1.42 and --0.24, respectively.
The colors of the BGCs and RGCs expected from the $B$-band luminosity of M85 are $(g-i)_0$ = 0.716 and 0.983, as mentioned in Section 3.4.
These are marked by a circle and a square in the figure.

The mean colors of the BGCs and RGCs are $(g-i)_0$ = 0.68 and 0.89, respectively, for the central region at $R < 2\arcmin$, and $(g-i)_0$ = 0.64 and 0.87, respectively, for the outer region at $4\arcmin < R < 6\arcmin$, as marked by dashed and dotted lines in the figure.
Therefore, the RGCs in the central region of M85 are $\sim$ 0.1 mag bluer than those for the GCs in the central region of other Virgo early-type galaxies, while the BGCs in the same region show a much smaller difference.
A comparison with simple stellar population models indicates that the RGC population in M85 might include \replaced{an 2-4 Gyrs old GCs}{intermediate-age GCs with ages of 2-4 Gyr and} higher metallicities than old GCs, or an old population with 
[Fe/H]$= -0.6$, which is much lower than the metallicities of the RGCs in the other Virgo early-type galaxies with similar luminosity.
The presence of 4 Gyr old GC populations in the central region of M85 found in the spectroscopic study \citep{ko18} supports the first possibility.
	
\section{Summary}

We presented the results from a wide-field photometric survey of GCs in \deleted{1 deg $\times$ 1 deg field around} M85 using CFHT/MegaCam.
We identified 1318 GC candidates with 20.0 mag $< g_0 <$ 23.5 mag in the survey region using concentration index, color, and magnitude criteria.
About 60\% of these are located inside $R = 20\arcmin$ from the galaxy center, and they are mostly GCs in M85.
The main photometric properties of the GCs are summarized as follows:

\begin{itemize}

\item
The spatial distribution of the GCs shows a strong concentration in the M85 center.
The radial number density profile of these GCs is {\bf well fit} by a S{\'e}rsic profile with $n = 2.58^{+0.43}_{-0.33}$.
The effective radius of the GC system is derived to be 4.14$^{+0.68}_{-0.51}$ arcmin, corresponding to 22$^{+4}_{-3}$ kpc.

\item
The $(g-i)_0$ color distribution of the GCs is not unimodal.
According to the GMM test for bimodality, the peaks of the color distribution are found at $(g-i)_0$ = 0.66 and 0.90. 
If we apply the same test for trimodality, we obtain $(g-i)_0$ = 0.64, 0.78, and 0.93.
However, the number fraction of the intermediate color component in this case is estimated to be very low with a value of $15^{+15}_{-10}$\%. Therefore, we conclude that the color distribution of the M85 GCs is bimodal. 
The intermediate-color GC candidates might exist only in the central region at $R < 4\arcmin$ based on the GMM results.

\item
The BGC system is more extended than the RGC system.
The effective radii of the BGC and RGC systems are 5.75$^{+0.91}_{-0.96}$ and 2.02$^{+0.47}_{-0.06}$ arcmin, respectively.

\item
The spatial distribution of the M85 GC system shows significant elongation, which is similar to that of \added{the} galaxy stellar light in the outer region ($R > 5\arcmin$). 
Interestingly, both BGC and RGC systems show a spatial correlation with \added{the} galaxy stellar light.
We suspect that an off-center major merger that occurred recently resulted in this elongated spatial distribution of the GC system and stars, as well as their rotation 
around the minor axis of M85 \citep{ko18}.

\item
We found two substructures of the GCs around M85, SUB-S and SUB-W, of which radial extents are about 13 kpc.
We estimated the total number of 
GCs in 
SUB-S and SUB-W to be 209 and 171, respectively.
These numbers indicate that their progenitors have a dynamical mass of $M_{\rm dyn} = 3-4 \times 10^{10} M_{\odot}$.

\item
The number fraction of the RGCs ($\sim$ 36-41\%) in the central region of M85 is much smaller than that in the other Virgo early-type galaxies that have luminosities similar to that of M85
($\sim$ 60\%).
This implies that the progenitor of M85 is less massive than that of massive early-type galaxies with similar luminosity.

\item
The mean $(g-i)_0$ color of the RGCs in M85 is found to be $\sim$ 0.1 mag bluer than those for other Virgo early-type galaxies with similar luminosity, 
while the mean color of the BGCs shows a small difference. 
This indicates that the RGCs might be a 2--4 Gyr old population, which is confirmed spectroscopically \citep{ko18}.

\end{itemize}

The properties of the M85 GC system indicate that the formation process of M85 is different from typical massive early-type galaxies.
Further simulation studies on the formation of M85 are needed to explain the unique GC system of M85.

\acknowledgments
The authors would like to thank the referee for the useful suggestions that helped to improve the original manuscript.
The authors also thank Jisu Kang for reading the manuscript and Brian S. Cho for improving the English in the manuscript.
This work was supported by the National Research Foundation of Korea (NRF) grant
funded by the Korean Government (NRF-2017R1A2B4004632) and the K-GMT Science Program (PID: cfht\_2014\_00006) funded through the Korean GMT Project operated by the Korea Astronomy and Space Science Institute (KASI).

\clearpage

%
%

%

\end{document}